\documentclass[aps,showpacs,superscriptaddress,nofootinbib,floatfix]{revtex4}
\usepackage{graphicx}
\def\tstrut{\vrule height2.5ex depth0pt width0pt} 

\begin{document}

\title{Neutrino Induced Coherent Pion Production off Nuclei and PCAC}

\author{E. \surname{Hern\'andez}} 
\affiliation{Grupo de F\'\i sica Nuclear, Departamento
de F\'\i sica Fundamental e IUFFyM,\\ Facultad de Ciencias, E-37008
Salamanca, Spain.}  
\author{J.  \surname{Nieves}}
\affiliation{ Instituto de F\'\i sica Corpuscular (IFIC), Centro Mixto
  CSIC-Universidad de Valencia, Institutos de Investigaci\'on de
  Paterna, Aptd. 22085, E-46071 Valencia, Spain} 
\author{M.J. \surname{Vicente Vacas} }
\affiliation{Departamento de F\'\i sica Te\'orica and IFIC, Centro Mixto
  CSIC-Universidad de Valencia, Institutos de Investigaci\'on de
  Paterna, Aptd. 22085, E-46071 Valencia, Spain}
  

\today
\begin{abstract}
We  review   the Rein--Sehgal model and criticize its use for
low energy neutrino induced coherent pion production. We have 
studied the validity of the main approximations implicit in  that
model, trying to compare with physical observables when that is
possible and with microscopical calculations. Next, we have tried to
elaborate a new improved model by removing the more problematic
approximations, while keeping the model still reasonably simple. Last,
we have discussed the limitations intrinsic to any approach based on
the partial conservation of the axial current hypothesis. In
particular, we have shown the inability of such models to determine the angular
distribution of the outgoing pion with respect to the direction of the
incoming neutrino, except for the $q^2= 0$ kinematical
point.
\end{abstract}

\maketitle

\section{Introduction}

The theoretical modelling of neutrino-induced coherent pion production in nuclei 
 in the low ($\sim$ 1 GeV) energy region is a quite
complicated task. The reasons are twofold. First, our limited
knowledge of the weak pion production on nucleons at these energies,
(see for instance Refs.~\cite{Llewellyn Smith:1971zm,Adler:1975mt,Fogli:1979cz,AlvarezRuso:1997jr, AlvarezRuso:1998hi,Slaughter:1998cw,Paschos:2000be,Lalakulich:2005cs,Hernandez:2007qq}),
that relies upon conflicting and low statistics bubble chamber
experimental data~\cite{Radecky:1981fn,Kitagaki:1986ct}.  Second, the
need of a quantum mechanical treatment of the multiple scattering
involving the strong interaction of pions and nucleons that can hardly
be accommodated in typical MonteCarlo approaches.  Many different
models have been proposed. Some of them are based on the assumption
that coherent pion production is dominated by the divergence of the
axial current and the use of the Partial Conservation of the Axial
Current
(PCAC) hypothesis~\cite{Rein:1982pf,Belkov:1986hn,Kopeliovich:2004px,
Gershtein:1980vd,Komachenko:1983jv,Paschos:2005km,Berger:2008xs,
Paschos:2009ag}. Other works develop a microscopical model for both
pion production and distortion assuming that the pion nuclear interaction
is dominated by the $\Delta(1232)$ resonance~\cite{Kim:1996az,Kelkar:1996iv,Singh:2006bm,AlvarezRuso:2007tt,AlvarezRuso:2007it,Amaro:2008hd}.
 Thus, they can only be applied at relatively low neutrino energies. 
Despite
the large effort dedicated to this process the situation is hardly
satisfactory and large discrepancies are found among different theoretical
predictions, both for  total and differential cross sections.

Besides, the understanding of these processes is very important for
the analysis of neutrino oscillation experiments. For instance charged
current (CC) coherent pion production is one of the candidate
processes responsible for the deficit found by the K2K collaboration
in the forward scattering events, which seriously limited the
prediction accuracy of the neutrino energy spectrum at the far
detector~\cite{Hasegawa:2005td}.  Similarly, neutral current (NC)
$\pi^0$ production is one of the largest background sources in muon
neutrino experiments, in particular for electron neutrino appearance
measurements \cite{AguilarArevalo:2008xs,AguilarArevalo:2007it}, 
because it can mimic electron events when one of the $\pi^0$ decay
photons is not detected. A significant part of the $\pi^0$'s in the
forward direction could come from coherent production.

Due to its success in the description of CC coherent pion production data at
high energies and also its simplicity and easy implementation, the Rein-Sehgal (RS)
model of Ref.~\cite{Rein:1982pf}  has been widely used by
 experimental collaborations  in their analyses
\cite{Hasegawa:2005td,AguilarArevalo:2008xs,Hiraide:2008eu}, even at
quite low neutrino energies clearly beyond the scope of the original
paper. We should remark that some of the RS model basic approximations
are better suited for high energies and heavy nuclei, as this implies
more forward peaked cross sections. In fact, the minimum neutrino
energy discussed in Ref.~\cite{Rein:1982pf} was 2 GeV whereas the
model has been used for the analysis of the coherent process at
neutrino energies down to the pion production threshold.

When compared to other approaches, which are more reliable for low
energy neutrinos, the differences clearly show up in the total
production rate (see e.g. Fig. 2 of Ref.~\cite{Hiraide:2008eu}) and
also in the pion angular or energy distributions (see Fig. 8 of
Ref.~\cite{Amaro:2008hd}\footnote{The curves labeled as MiniBooNE
Coll. correspond to the MiniBooNE implementation of the RS model}).
Some of the discrepancies, like the
wider angular distributions of the pions predicted by the RS model, can
be easily traced back to the approximations used in its derivation and
they could lead to important consequences in the determination of the
size of the pion yield and/or the ratio of coherent to non-coherent
pion production for low energy incident neutrinos.

Given the deficiencies of the RS model for low neutrino energies, it
seems clear the need of the use of alternative approaches. However, up
to our knowledge, the currently available models are of limited value
for that purpose. On the one hand, the PCAC based calculations share
many of the limitations of the RS model that will be discussed
below. On the other hand, the microscopical models based on the
$\Delta$ dominance are expected to be reliable only for low energy
neutrinos ($\lesssim$ 1 GeV).

Some important steps have been given in the direction of developing a
model for pion production on the nucleon beyond the $\Delta$
resonance, e.g. Ref.~\cite{Lalakulich:2006sw}, but we are still in need of
more experimental data (cross sections) that could better constrain
the little known transition form factors.  This is a long term project
that will require a collaborative effort of both experimental and
theoretical groups.

In the meantime, and lacking a microscopical approach for the whole
energy region of interest, it looks worthwhile to try to modify the RS
model to extend its applicability towards lower neutrino energies. Some
recent works in this line include
Refs.~\cite{Rein:2006di,Berger:2008xs} which already correct some of
the known RS problems.  In this work, we will further explore this
possibility. First, we will carefully study the validity of the main
approximations implicit in the RS model, trying to compare with
physical observables when that is possible and with microscopical
calculations. Second, we will try to elaborate a new improved model by
removing the more problematic approximations, while keeping the model
still reasonably simple. Last, we will consider the limitations
intrinsic to any PCAC based approach.

The paper is organized as follows. After some basic formalism in
Sect.~\ref{sec:sf}, we review the RS model
in Sect.~\ref{sec:rs}. In Sect.~\ref{sec:im} we present our improved
model. In Sect.~\ref{sec:pns} we confront an extension of both models with
low energy pion-nucleus scattering data. In Sect.~\ref{sec:re}, we show the results of
 both models for low energy neutrino-induced coherent pion production and
we also compare them  with a  microscopical
calculation that we consider reliable in that energy region. Finally,
we summarize the paper in Sect.~\ref{sec:su}.


\section{NC Neutrino Coherent Pion Production Structure functions}
\label{sec:sf}

We will focus on NC $\pi^0$ coherent production off a nucleus in its ground state ${\cal N}_{gs}$,
\begin{equation}
  \nu_l (k) + {\cal N}_{gs} \to \nu_l (k^\prime) + {\cal N}_{gs} +
  \pi^0(k_\pi)\, .
\label{eq:neureac}
\end{equation}
The modifications required for the CC case are straightforward.
The process starts with a weak pion ($\pi^0$) production followed by
the strong distortion of the pion in its way out of the nucleus. The
nucleus is left in its ground state unlike the case for incoherent
production where the nucleus is either broken or left in some excited
state.

Defining the four momentum transfer $q=k-k^\prime$ and taking
$\vec{q}$  and $\vec k\times\vec k\,'$along
the positive $z-$ and  $y-$axis respectively, one can write the differential cross section
with respect to the outgoing neutrino variables, the Mandelstam
variable $t=(q-k_\pi)^2$
and the pion azimuthal angle, $\phi_{k_\pi  q}$, in the LAB
system as:
\begin{eqnarray}
 \frac{d\sigma}{dE' d\Omega(\hat{k'}) dt\, d\phi_{k_\pi  q}} =  
\frac{|\vec{k}'|}{|\vec{k}|}\frac{G^2}{16\pi^2}
 L_{\mu\sigma } {\cal H}^{\mu\sigma} \,,
\label{eq:sig-gen}
\end{eqnarray}
with $E'$ the energy of the final neutrino and $G$ the Fermi
constant. The leptonic tensor is given by:
\begin{eqnarray}
{ L}_{\mu\sigma}
=
 k^\prime_\mu k_\sigma +k^\prime_\sigma k_\mu
+  \frac{q^2}{2} g_{\mu\sigma} + {\rm i}
\epsilon_{\mu\sigma\alpha\beta}k^{\prime\alpha}k^\beta \,,
\label{eq:lep}
\end{eqnarray}
and it is orthogonal to the transferred four momentum $q^\mu$. In our
convention, we take $\epsilon_{0123}= +1$ and the metric
$g^{\mu\nu}=(+,-,-,-)$. The hadronic part, ${\cal H}$, is determined
by the matrix element of the NC between the initial and final hadronic
states, and it includes all the nuclear effects.
Introducing the variable $y=q^0/E$, with $E$  the incident neutrino energy, we
can write
\begin{equation}
\frac{d\sigma}{dq^2dydt\, d\phi_{k_\pi  q}} = \frac{G^2}{16\pi^2} E\, \kappa
\left(-\frac{q^2}{|\vec{q }\,|^2}\right)\, (u^2\frac{d
  \sigma_L}{dt\, d\phi_{k_\pi  q}}+v^2\frac{d\sigma_R}{dt\, d\phi_{k_\pi  q}}
+2 u v\frac{d\sigma_S}{dt\, d\phi_{k_\pi  q}}+\frac{d{\cal A}}{dt\, d\phi_{k_\pi q}})  \,,
\label{eq:dsigmarsl}
\end{equation}
where  
\begin{equation}
\kappa=q^0+\frac{q^2}{2{\cal M}}, \quad\;\; 
u,v=\frac{E+E'\pm |\vec{q}\,|}{2E}\,,
\end{equation}
and
\begin{eqnarray}
\label{eq:dsigma_S}
\frac{d\sigma_S}{dt\, d\phi_{k_\pi  q}}&=&-\frac{1}{q^2}\,
\frac{\pi}{\kappa}\left(
|\vec q\,|^2 {\cal H}_{00}+q^0|\vec q\,|\,({\cal H}_{0z}+{\cal
  H}_{z0})+ (q^{0})^2
 \,{\cal H}_{zz}\right)\,,\nonumber\\
\frac{d\sigma_L}{dt\, d\phi_{k_\pi  q}}&=&
\frac{\pi}{2\kappa}\left({\cal H}_{xx}+{\cal H}_{yy}
+i\,({\cal H}_{xy}-{\cal H}_{yx})\right)
\,,\nonumber\\
\frac{d\sigma_R}{dt\, d\phi_{k_\pi  q}}&=&
\frac{\pi}{2\kappa}\left({\cal H}_{xx}+{\cal H}_{yy}
-i\,({\cal H}_{xy}-{\cal H}_{yx})\right)
 \,,\nonumber\\
 \frac{d{\cal A}}{dt\, d\phi_{k_\pi q}}&=&\frac{\pi}{\kappa}
\left(uv\,({\cal H}_{xx}-{\cal H}_{yy})
+\frac{E+E\,'}{E}\sqrt{\frac{|\vec{q}\,|^2}{-q^2}} \sqrt{uv}\,\left((
{\cal H}_{0x}+{\cal H}_{x0})+\frac{q^0}{|\vec{q}\,|}\,(
{\cal H}_{zx}+{\cal H}_{xz})\right)\right.\nonumber\\
&&+\left.
i\frac{|\vec q\,|}{E}\sqrt{\frac{|\vec{q}\,|^2}{-q^2}} \sqrt{uv}\,\left((
{\cal H}_{0y}-{\cal H}_{y0})+\frac{q^0}{|\vec{q}\,|}\,(
{\cal H}_{zy}-{\cal H}_{yz})\right)\right)\,.
\end{eqnarray}
Note that
\begin{eqnarray}
\frac{d\sigma_S}{dt\, d\phi_{k_\pi  q}}=\frac{1}{2\pi}\,\frac{d\sigma_S}{dt}\
\ , \ \ 
\frac{d\sigma_R}{dt\, d\phi_{k_\pi  q}}=\frac{1}{2\pi}\,\frac{d\sigma_R}{dt}\
\ , \ \ 
\frac{d\sigma_L}{dt\, d\phi_{k_\pi  q}}=\frac{1}{2\pi}\,\frac{d\sigma_L}{dt}\,,
\end{eqnarray}
as neither ${\cal H}_{00}$, ${\cal H}_{zz}$ nor the combinations
${\cal H}_{0z}+{\cal H}_{z0}$, ${\cal H}_{xx}+{\cal H}_{yy}$, ${\cal
H}_{xy}-{\cal H}_{yx}$ can depend on $\phi_{k_\pi q}$. The only
quantity that depends on $\phi_{k_\pi q}$ is $\frac{d{\cal
A}}{dt\,d\phi_{k_\pi q}}$. The latter is not a proper differential
cross section as it can take on positive as well as negative
values. Besides, it cancels upon integration on $\phi_{k_\pi q}$ (see
below). Note also that for $q^2=0$ only the $\sigma_S$ term
contributes.

The tensor $H^{\mu\nu}$,
\begin{equation}
H^{\mu\nu}=\int\,d\phi_{k_\pi q}\,{\cal H}^{\mu\nu} \,,
\end{equation}
only depends on $q^\mu$, $p^\mu= ({\cal M}, \vec{0})$ (the four vector
of the initial nuclear state, with ${\cal M}$ the nucleus mass) and
$t$.  Due to the tensorial character of $H$ and the fact that, being
the transferred momentum $\vec{q}$ aligned with the $z-$axis, it is
invariant under rotations around the $z-$axis, one can prove
\begin{equation}
H^{xx}= H^{yy}, \quad  H^{xz}= H^{zx}=H^{yz}= H^{zy}=H^{0x}=
H^{0y}=H^{x0}= H^{y0}=0, \quad H^{xy}= - H^{yx}    \,,
\label{eq:sym}
\end{equation}
Thus, integrating in $\phi_{k_\pi q}$ 
one obtains for the $\frac{d\sigma}{dq^2dydt}$ differential cross 
section~\cite{Lee:1962jm}
\begin{equation}
\frac{d\sigma}{dq^2dydt} = \frac{G^2}{16\pi^2} E\, \kappa
\left(-\frac{q^2}{|\vec{q }\,|^2}\right)\, (u^2\frac{d
  \sigma_L}{dt}+v^2\frac{d\sigma_R}{dt}
+2 u v\frac{d\sigma_S}{dt})  \,,
\label{eq:sigmarsl}
\end{equation}
where  
\begin{eqnarray}
\label{eq:sigma_S}
\frac{d\sigma_S}{dt}&=&-\frac{1}{q^2}\,\frac{\pi}{\kappa}\left(
|\vec q\,|^2 H_{00}+q^0|\vec q\,|\,(H_{0z}+H_{z0})+ (q^{0})^2
 \,H_{zz}\right)\,,\nonumber\\
\frac{d\sigma_L}{dt}&=&\frac{\pi}{\kappa}\left(H_{xx}+i\,H_{xy}\right)
\,,\nonumber\\
\frac{d\sigma_R}{dt}&=&\frac{\pi}{\kappa}\left(H_{xx}-i\,H_{xy}\right)
 \,.
\end{eqnarray}
As stated above, at $q^2=0$ only $\sigma_S$ contributes, 
and given that  in this case
$q^0=|\vec{q}\,|$, one  finds that  the cross section goes as
\begin{equation}
\left(
|\vec q\,|^2 H_{00}+q^0|\vec q\,|\,(H_{0z}+H_{z0})+ (q^0)^2
 \,H_{zz}\right)
 = q^\mu q^\nu H_{\mu\nu} \,.
\end{equation}
In other words, in the $q^2=0$ limit, the lepton tensor of
Eq.~(\ref{eq:lep}), turns out to be proportional to $q^\mu q^\nu$ and
thus, one is left to compute the matrix element of the divergence of
the hadronic current. Since the vector NC is conserved, only the
divergence of the axial part contributes to the cross section.  PCAC
can then be used to express the divergence of the axial current in
terms of the pion field operator $\Phi(x)$,
\begin{equation}
 \partial_\mu A^\mu_{NC}(x) = 2 f_\pi m_\pi^2 \Phi(x)\,,
\label{eq:pcac-ok}
\end{equation}
with $f_\pi \approx 92.4$ MeV, the pion decay constant. Treating the
nucleus as an elementary particle, one can write
\begin{equation}
\left \langle {\cal N}_{gs} \pi^0 (k_\pi) | q_\mu A^\mu_{NC} |{\cal
  N}_{gs} \right \rangle_{q^2=0} = -2{\rm i} f_\pi T\left(
  {\cal N}_{gs} \pi^0 (k_\pi) \leftarrow \pi^0 (q) {\cal N}_{gs}
  \right) \Big|_{q^2=0}\,,
\label{eq:pcac}
\end{equation}
where $T (f \leftarrow i)$ is the transition
amplitude  between the initial hadron state plus a pion of four
momentum $q^\mu$, and the final hadronic state. Using this
relation, one obtains
\begin{equation}
\left.q^2\,\frac{d\sigma_S}{dt}\right|_{q^2= 0}=-\left.4
\frac{E_\pi}{\kappa}\,f_\pi^2\,
\frac{d\sigma(\pi^0 {\cal N}_{gs} \to \pi^0 {\cal N}_{gs})}{dt}\right|_{q^2=0}\,,
\label{eq:rs07}
\end{equation}
and then, neglecting the nucleus recoil $(q^0=E_\pi)$, one can further
write (Adler's PCAC formula~\cite{Adler:1964yx})
\begin{equation}
\left.\frac{d\sigma}{dq^2dydt}\right|_{q^2=0} = \frac{G^2f_\pi^2}{2\pi^2}
\frac{E\,u\,v}{|\vec q\,|}
\left.\frac{d\sigma(\pi^0 {\cal N}_{gs} 
\to \pi^0 {\cal N}_{gs})}{dt}\right|_{q^2=0,\,E_\pi=q^0}\,.
\label{eq:rs08}
\end{equation}
In the  $q^2=0$ limit that we are using
\begin{equation}
\frac{Euv}{|\vec q\,|}=\frac{1-y}{y},
\end{equation}
and thus 
\begin{equation}
\left. \frac{d\sigma}{dx\,dy\,dt}\right|_{q^2=0} = \frac{G^2 M
E}{\pi^2}f_\pi^2 (1-y) \frac{d\sigma(\pi^0 {\cal N}_{gs} \to \pi^0
{\cal N}_{gs})}{dt}\Big|_{q^2=0,\,E_\pi=q^0}\,,
\label{eq:rsNucleo}
\end{equation}
with $x=-q^2/2Mq^0$ and $M$ the nucleon mass. This latter form,
Eq.~(\ref{eq:rsNucleo}), was adopted in the original RS
model~\cite{Rein:1982pf}.  Recently, Berger and
Sehgal~\cite{Berger:2008xs} have proposed to use Eq.~(\ref{eq:rs08})
instead, for finite $q^2$. To our understanding, both choices are
somehow arbitrary. As shown in the derivation, there are several
factors of the type $q^0/|\vec{q}\,|$, which do not affect the $q^2=0$
calculation, and that might lead to different corrections far from
that limit.

An important remark is in order here. Adler's PCAC formula relates the
neutrino induced cross section to the off-shell ($0=q^2\ne
m_\pi^2=k_\pi^2$) elastic pion--nucleus one (see Eqs.~(\ref{eq:rs07}),
(\ref{eq:rs08}), and (\ref{eq:rsNucleo})). It is tempting to neglect
off--shell effects and approximate the latter cross section by the
experimental one, or by any realistic model for it. This can be only
strictly correct in the case of a pointlike nucleus. This was firstly
pointed out by J.S. Bell shortly after Adler proposed his PCAC
formula. Because of absorption and inelastic collisions, physical
pions do not penetrate into the interior of heavy nuclei, and thus the
pion-nucleus elastic cross section is only sensitive to the nuclear
surface (roughly it scales as $A^\frac23$, with $A$ the nucleus mass
number). But neutrinos penetrate to all parts of nuclei, being then
sensitive to the whole nuclear volume; for them cross sections scale
as $A$~\cite{Bell:1964eu}. Thus Bell expected, that although the
on-shell $\sigma_{\pi^0{\cal N}}$ cross section is surface-like, the
off-shell ($q^2=0$) one turned out to be volume-like. That clearly
hints at a non-trivial off-shell behaviour for $\sigma_{\pi^0{\cal N}}$.
Similar arguments were raised and discussed by Bernabeu, Ericson and
Jarlskog in the context of muon capture in nuclei~\cite{Bernabeu:1977tt}.

With all these caveats, there are cases where approximating the off-shell
pion-nucleus cross section by the on-shell one in 
Eqs.~(\ref{eq:rs07}), (\ref{eq:rs08}), and (\ref{eq:rsNucleo}) could
be accurate and easily obtained in any good model for pion nuclear
scattering. For instance, i) long wavelength limit: the pion
wavelength is larger than the nucleus and it probes the whole nuclear
volume (this in the case of the nuclear beta decay), or ii)
short wavelength limit: here the pion interaction is weak and there is
little multiple scattering. Practically we have a collection of
independent scatterers (nucleons).

This problem has not been taken into account in the recent works of
Refs.~\cite{Berger:2008xs,Paschos:2009ag}, where it is proposed to use
the experimental elastic pion-nucleus cross section in the Adler's
PCAC formula at energies where the process is dominated by the weak
excitation of the $\Delta(1232)$ resonance and its subsequent decay
into a $\pi N$ pair. For resonant energies, the pion wavelength is
such that it renders doubtful the assumption of a pointlike nucleus,
while the pion--nucleon interaction is sufficiently strong to expect
that surface effects due to the distortion of the incoming pion waves,
included in the experimental elastic pion-nucleus cross section, might
induce  inaccuracies in the computation of the weak process.

In order to remove such an effect from the physical
$\sigma_{\pi^0{\cal N}}$ one should rely on a distortion model, for
instance the one described below in Subsect.~\ref{sec:out-dist}. The
original model of Ref.~\cite{Rein:1982pf} deals with this {\it
volume-surface} problem when $<x>$, the average path length of the
pions, is calculated by choosing the path length traversed by pions
uniformly produced in the nucleus instead of the path length of pions
scattered on the nucleus.

\section{Review of the Rein--Sehgal model}
\label{sec:rs}

As discussed above, the Rein--Sehgal model~\cite{Rein:1982pf} uses the
Adler's PCAC formula~\cite{Adler:1964yx} and approximates the coherent
$\pi^0$ production differential cross section in the laboratory frame
by means of Eq.~(\ref{eq:rsNucleo}).  Besides the neglect of the
off--shell $(q^2\ne m_\pi^2)$ effects in the right hand side of
Eq.~(\ref{eq:rsNucleo}), this expression already involves a few
approximations, that will not be discussed in this work. For instance,
the final nucleus recoil energy is also neglected, which allows to
approximate $q^0$ by the pion energy $E_\pi$. 
 
In Ref.~\cite{Rein:1982pf}, the expression of Eq.~(\ref{eq:rsNucleo})
for neutrino coherent pion production is continued to non--forward
lepton directions $(q^2\ne 0)$ by attaching a propagator term
$1/(1-q^2/m_A^2)^2$ with $m_A\approx 1$ GeV. This does not mean that
the $q^2$ dependence is merely given by the added propagator term, and
in fact it is much more pronounced than that induced by this
factor. This is because large (and negative) $q^2$ values are
suppressed by the elastic pion--nucleus differential cross section
that strongly favours $t=0$. Within the $q^0=E_\pi$ approximation, a
zero value for $t$ would imply $q^2= m_\pi^2$, a kinematical point
that cannot be reached since $q^2$ is space-like.  Thus, the lowest
values accessible for $t$ must come  from the $q^2 \approx 0$
region.

Once $q^2\ne 0$ there are other finite contributions stemming from
$\sigma_L$, $\sigma_R$ and additional pieces of $\sigma_S$ fully
disregarded in the RS model.  These contributions come from both the
axial and the vector part of the NC.  These further approximations
have been, somehow, justified because of the strongly forward peaked
character of the process that implies that only quite small values of
$q^2$ will be relevant. This assumption will be discussed later in
more detail.  In particular, as the contribution of the vector current
is neglected, the model predicts equal neutrino and antineutrino cross
sections.

Next, in Ref.~\cite{Rein:1982pf}, the pion--nucleus cross section,
with the caveats mentioned above, is expressed in terms of the
pion--nucleon one and then one obtains
\begin{equation}
\frac{d\sigma}{dx\,dy\, dt}
= \frac{G^2 M
E}{\pi^2}f_\pi^2 (1-y)\frac{1}{(1-q^2/1\ {\rm GeV}^2)^2} 
\left (|F_{\cal A}(t)|^2 F_{\rm abs}
\frac{d\sigma(\pi^0 N \to \pi^0  N)}{dt}\Big|_{E_\pi=q^0, t=0}\right )\,,
\label{eq:rsNucleon}
\end{equation}
where $F_{\cal A}(t)$ is the nuclear form factor which can be
calculated as $F_{\cal A}(t)=\int d^3\vec{r}\ e^{{\rm
i}\left(\vec{q}-\vec{k}_\pi\right)\cdot\vec{r}}
\left\{\rho_p(\vec{r}\,)+\rho_n(\vec{r}\,)\right \}$, with the density
$\rho_{p(n)}$ normalized to the number of protons (neutrons).
Finally, according to~\cite{Rein:1982pf}, $F_{\rm abs}$ is a
$t-$independent attenuation factor that takes into account effects of
the outgoing pion absorption in the nucleus\footnote{As defined in
~\cite{Rein:1982pf}, $F_{\rm abs}$ only removes from the flux pions
that undergo inelastic collisions but no true pion absorption is
actually accounted for by means of this factor~\cite{Amaro:2008hd}.}.
 
At this point, some new approximations have been implemented to
estimate the pion-nucleus cross section.  First, the pion--nucleon
cross section is evaluated at $t=0$, namely, in the forward
direction. This can be justified if the nuclear form factor is
sufficiently forward peaked. 
The larger the pion energy and the
heavier the nucleus, the better this approximation becomes. In the
original paper~\cite{Rein:1982pf}, Eq.~(\ref{eq:rsNucleon}) was
employed for a medium size nucleus, aluminum, and neutrino energies
above 2 GeV, for which the relevant pion energies are quite
high. However, as we will show below and it was already pointed out in
~\cite{Amaro:2008hd}, for neutrino energies below 1 GeV and lighter
nuclei, like carbon or oxygen, the nuclear form factor is not enough
forward peaked to render the finite $t-$dependence of the
pion--nucleon cross section negligible, and even in the forward direction the
$t$ value is not  close enough to zero. Second, the distortion factor $F_{\rm
abs}$ is an oversimplification since in any realistic scattering model
this factor should depend on $t$.

\section{Improvements on the RS model}
\label{sec:im}
 
There exist sophisticated microscopical calculations of the neutrino
coherent $\pi^0$ production off nuclei based on the dominance of the
$\Delta(1232)$ resonance~\cite{Kim:1996az,Kelkar:1996iv,Amaro:2008hd,
Singh:2006bm,AlvarezRuso:2007tt, AlvarezRuso:2007it}.  However, these
are difficult to extend for the higher neutrino energies present in
the current experiments and to implement in Monte Carlo
algorithms. Actually, one of the virtues of the RS model is its
simplicity, that allows its use for different nuclei and pion
energies. This is one of the reasons why the RS model has been widely
used. 


We propose here a minimal extension
of the model in which we
implement two main corrections. Firstly, we will take into
account the $t-$dependence of the pion-nucleon cross section when
computing the pion-nucleus scattering, and secondly we will work out a
more realistic description of the outgoing pion
distortion\footnote{The pion-nucleus cross section can not be
factorized, in addition to the $\pi$ N elementary cross section, as
the product of a nuclear form factor (defined as the Fourier transform
of the density) and a distortion factor. This is in contrast with
purely electroweak processes not involving strong final state
interactions.}.

\subsection{$t$-dependence of the pion-nucleon cross section}

\begin{figure}[tbh]
\centerline{\includegraphics[height=10cm]{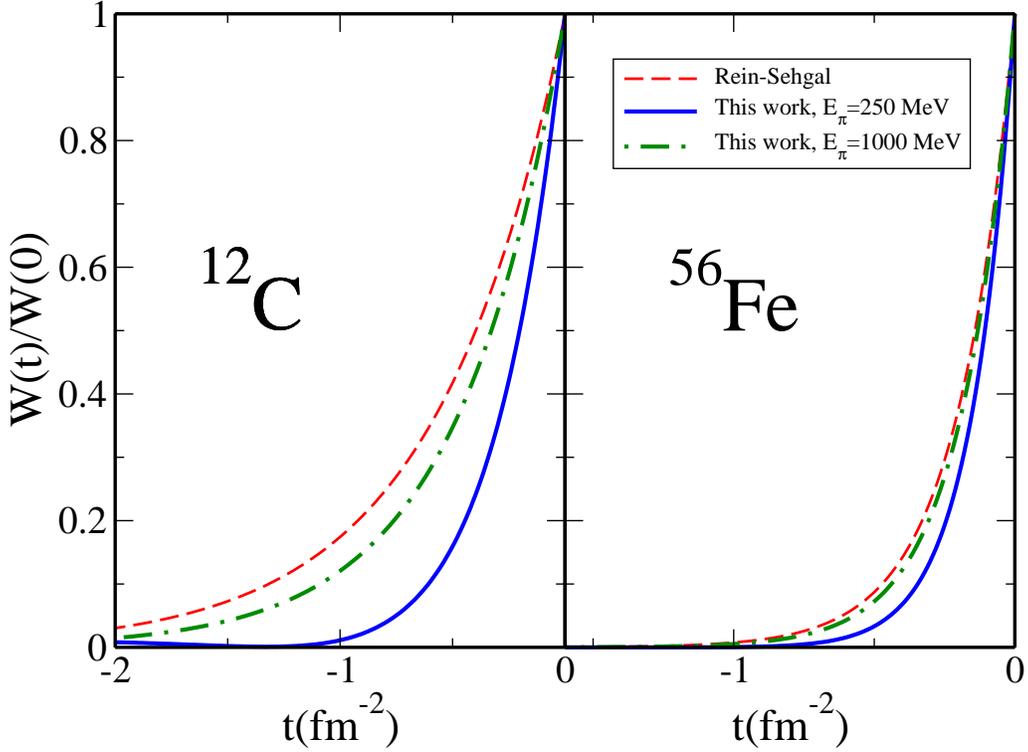}}
\caption{$W(t)/W(0)$ ratio for carbon and iron at two pion energies. 
 }\label{fig:1NM}
\end{figure}

The $t=0$ approximation in the right hand side of
Eq.~(\ref{eq:rsNucleon}) is unnecessary and can be easily removed, and
one obtains
\begin{equation}
\left( \frac{d\sigma}{dx\,dy\, dt}\right)_{q^2=0} = \frac{G^2 M
E}{\pi^2}f_\pi^2 (1-y) |F_{\cal A}(t)|^2 F_{\rm abs}
 \frac{d\sigma_{nsf}(\pi^0 N \to \pi^0  N)}{dt}\Big|_{E_\pi=q^0}\,,
 \end{equation}
where $\sigma_{nsf}$ is the non spin-flip part of the pion nucleon
cross section because spin-flip processes do not contribute to the
elastic pion nucleus cross section. Given that for $t=0$ the spin-flip
cross section is identically zero this constraint was unnecessary in
Ref.~\cite{Rein:1982pf}.  The non spin--flip cross section can be
evaluated in terms of the phase shifts ($\delta^J_{I,\ell}$) and
inelasticities ($\eta^J_{I,\ell}$), that we take from
Ref.~\cite{Arndt:1995bj} ($J,\,\ell$ and $I$ stand respectively for
total angular momentum, orbital angular momentum and isospin of the
pion-nucleon pair). For an incoming neutral pion of laboratory energy $E_\pi$,
the non spin-flip cross section reads
\begin{eqnarray}
\frac{d\sigma_{nsf}}{dt} &=& 
\frac{\pi}{9 |\vec{k}_\pi|_{c.m.}^2}\Big | 2 a^{\frac32}
+ a^\frac12 \Big |^2\,, \nonumber\\
a^I(s,t) & = & \sum_\ell \left[ (\ell +1 ) f_{I,\ell}^{\ell+1/2}
+ \ell \, f_{I,\ell}^{\ell-1/2} \right] P_\ell(\cos\theta)\,, \nonumber \\
  f_{I,\ell}^{J}(s) &=& \frac{\eta^J_{I,\ell}\, e^{2{\rm i}\, 
\delta_{I,\ell}^{J}}-1}{2{\rm i} |\vec{k}_\pi|_{c.m.}}, ~~ 
 \cos\theta = 1 + \frac{t}{2|\vec{k}_\pi|_{c.m.}^2}\,, \label{eq:coscm}
\end{eqnarray}
with $s=m_\pi^2+M^2+2ME_\pi$,
$|\vec{k}_\pi|_{c.m.}^2=\left(s-(M+m_\pi)^2\right)\left(s-(M-m_\pi)^2\right)
/4s$, the pion  momentum squared and  $P_\ell$  the Legendre
polynomials. For the sake of completeness we also give here the
expression for the inelastic pion-nucleon cross section, $\sigma_{inel}$
\begin{equation}
\sigma_{inel}(s) = \frac{4\pi}{|\vec{k}_\pi|_{c.m.}}{\rm Im}\left(
\frac23 a^{\frac32}(s,t=0)+ \frac13 a^\frac12 (s,t=0)\right)-4\pi
\sum_\ell \left[ (\ell +1 ) |a_{\ell}^{\ell+1/2}|^2+ \ell \,
  |a_{\ell}^{\ell-1/2}|^2 \right] \,,
\end{equation}
with $a_{\ell}^J(s)= \left(2\,f_{3/2,\ell}^J(s) +
f_{1/2,\ell}^J(s)\right)/3$.  
To discuss the importance of the $t$ dependence of the cross section  
we plot in Fig.~\ref{fig:1NM} the ratio $W(t)/W(0)$ for carbon and iron  at
two different pion energies . The function $W(t)$ is defined as
\begin{equation}
W(t)=|F_{\cal A}(t)|^2 \frac{d\sigma_{nsf}(\pi^0 N \to \pi^0  N)}{dt}\, ,
\label{eq:rsxxx}
\end{equation}
and for the RS model $\frac{d\sigma_{nsf}(\pi^0 N \to \pi^0 N)}{dt}$
is taken at $t=0$.  To allow for a better comparison with the RS model
we have used the same nuclear form factor as in
Ref.~\cite{Rein:1982pf}, given by 
$F_{\cal A}(t)=A\,e^{bt/6}$, with $b=R_0^2A^{2/3}$ and $A$ the number of
nucleons. As in Ref.~\cite{Rein:1982pf} we take $R_0=1\,$ fm.  As can be appreciated in
the figure, the extra dependence of the non-spin flip pion-nucleon
cross section on $t$, neglected in Ref.~\cite{Rein:1982pf}, produces
drastic changes in a light nucleus like $^{12}$C for low energy
pions. We have chosen $E_\pi=250$ MeV because it corresponds
approximately to the maximum of the $\pi^0$ yield in the MiniBooNE
experiment~\cite{AguilarArevalo:2008xs}. The corrected distribution is
much narrower and leads to a significantly smaller area. Effects
become less important for increasing pion energies and for heavier
nuclei thanks to the more pronounced behavior of the nuclear
form-factor. In the original work of Ref.~\cite{Rein:1982pf}, aluminum
and neutrino energies, at least of 2 GeV, were considered. For 2 GeV
neutrinos, the pion spectrum will peak around 1 GeV, and for such pion
energies the corrections to the RS formula turn out to be moderately
small. But as seen, they can be relevant for the low energy pions produced in 
low energy neutrino processes.

\subsection{Pion distortion}
\label{sec:out-dist}
Next, we pay attention to the distortion of the outgoing pion. We use
an eikonal approximation, in the spirit of the original RS model, and
taking $\vec{q}$ in the $z$ direction, we replace
\begin{equation}
|F_{\cal A}(t)|^2 F_{\rm abs} \equiv \Big | \int d^3\vec{r}\ e^{{\rm
i}\left(\vec{q}-\vec{k}_\pi\right)\cdot\vec{r}} \rho(\vec{r})\Big|^2 F_{\rm abs}\end{equation}
by 
\begin{eqnarray}
|F_{\cal A}^{distor}(|\vec{q}\,|,|\vec{k}_\pi\,|,
\vec{q}\cdot\vec{k}_\pi)|^2 &\equiv& \Big | \int d^3\vec{r}\ e^{{\rm
    i}\left(\vec{q}-\vec{k}_\pi\right)\cdot\vec{r}} \rho(\vec{r}\,)
\Gamma(b,z) \Big |^2 \\ &=& \Big | 2\pi\int_0^{+\infty} db\, b\,
J_0(b\, p_T)
\int_{-\infty}^{+\infty}dz\,\rho(\sqrt{b^2+z^2})\Gamma(b,z) e^{{\rm
    i}p_z z}\Big |^2\,,
\end{eqnarray}
with $\rho(\vec{r}\,)$ the nuclear density, $b= \sqrt{x^2+y^2}$ the
 impact parameter,
 $p_z=|\vec{q}\,|-\vec{q}\cdot\vec{k}_\pi/|\vec{q}\,|$,
 $p_T=\sqrt{\vec{k}_\pi^2-(\vec{q}\cdot\vec{k}_\pi)^2/\vec{q}{\,^2} }$ 
and $J_0$ the Bessel function.  The eikonal
 distortion factor $\Gamma(b,z)$ is defined as
\begin{equation}
\Gamma(b,z) = \exp\left(
-\frac12 \sigma_{inel}\int_z^{+\infty} dz' \rho\left(
\sqrt{b^2+z^{\prime 2}} \right)\right)\,. \label{eq:gamma}
\end{equation}
If one takes $\Gamma(b,z)$ independent of $b$ and $z$, 
$F_{\cal A}^{distor}(|\vec{q}\,|,|\vec{k}_\pi\,|,
\vec{q}\cdot\vec{k}_\pi)$ depends only on $p_z^2+p_T^2=-t$ and  
it reduces to the structure originally proposed in
Ref.~\cite{Rein:1982pf}, where all $t$ dependence is ascribed to the
nuclear form factor $F_{\cal A}(t)$. Indeed, in
the original RS model some volumetric average for $\Gamma(b,z)$ is
assumed. Neglecting the $t-$dependence, inherited from the impact
parameter dependence of the distortion of the pion waves 
leads to flatter angular distributions as we will show.


\section{Pion Nucleus Scattering}
\label{sec:pns}
In this section we will confront with data the predictions of the two previous approaches
when extended to elastic pion nucleus scattering.  By studying
low energy pion-nucleus scattering we magnify the effect of the distortion factor 
and in this way we expect to gain 
some insight into its possible deficiencies when applied to
low energy neutrino--induced coherent pion production.

 Following Ref.~\cite{Rein:1982pf}, one would write for the pion
nucleus elastic cross section
\begin{equation}
\frac{d\sigma(\pi^0 {\cal N}_{gs}
 \to \pi^0 {\cal N}_{gs})}{dt} = 
|F_{\cal A}(t)|^2 F^\pi_{\rm abs}
\frac{d\sigma(\pi^0 N \to \pi^0  N)}{dt}\Big|_{ t=0}\,,
\label{eq:rsPiNucleo}
\end{equation}
where we have replaced the original absorption factor $F_{\rm abs}$,
appropriate for neutrino induced pion production, by $F^\pi_{\rm abs}$
that also takes into account the initial pion distortion. We have done this
following Ref.~\cite{Rein:1982pf}. Namely, 
the original absorption factor $F_{\rm
abs}$ given in Ref.~\cite{Rein:1982pf} reads:
\begin{equation}
F_{\rm abs}=e^{-<x>/\lambda},
\end{equation}
where $<x>$ is the average path length traversed by a $\pi^0$ produced
homogeneously in the nuclear volume by the neutrino, and
$\lambda^{-1}=\sigma_{inel}\rho$ with $\sigma_{inel}$ the inelastic
$\pi^0$-nucleon cross section.  The average $<x>$ is found to be
$3R/4$ assuming a hard sphere density of radius $R$ for the
nucleus. Thus, taking $R=R_0 A^\frac13$, one finds
\begin{equation}
F_{\rm abs}=e^{-\frac{9A^{1/3}}{16\pi R_0^2}\sigma_{inel}},
\end{equation}
To properly compare with pion
nucleus data one should also consider the distortion of the incoming
pion. In this case, after averaging over the impact parameter one
obtains $<x>=4R/3 $ and thus we replace $F_{\rm abs}$ by $F_{\rm
abs}^{\pi}$ where
\begin{equation}
F_{\rm abs}^{\pi}=e^{-\frac{A^{1/3}}{\pi R_0^2} \sigma_{inel}}. \label{eq:fabs-pion}
\end{equation}
Using the original $F_{\rm abs}$ instead of $F_{\rm abs}^{\pi}$ amounts to 
a change of scale but does not modify the angular shape of the cross section.
 Finally, we obtain $\sigma_{inel}$ and
$\frac{d\sigma(\pi^0 N \to \pi^0 N)}{dt}\Big|_{ t=0}$ from the SAID
partial wave analysis \cite{Arndt:1995bj}; detailed expressions were
given before.
%
%
Before discussing the results, we would like to point out that the
calculation is for $\pi^0$-nucleus scattering while data correspond to
charged pions. However, we have only selected data from
isoscalar nuclei and except for Coulomb effects (relevant at small
angles) the cross section should be the same. 

We would like to stress the fact that the authors of Ref.~\cite{Rein:1982pf}
did not actually make a model for low and medium energy pion-nucleus scattering. The extension 
to those energies that we are showing here should certainly not be attributed to 
the RS model.

\begin{figure}[tbh]
\centerline{\includegraphics[height=10cm]{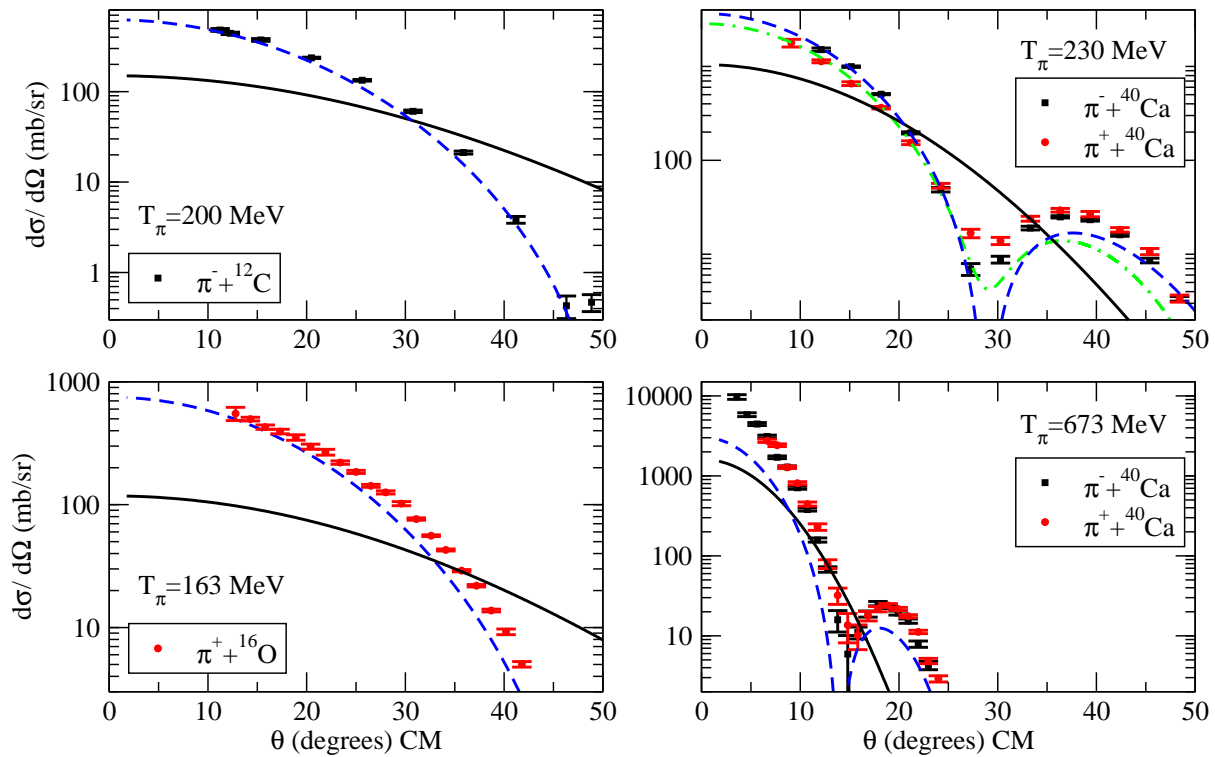}}
\caption{Pion nucleus elastic cross section for several nuclei and
  energies. Solid line: RS based model. Dashed line: Improved model of
  Eq.~(\ref{eq:piN}). For comparison in the right top panel, we also
  display results (dashed dotted green line) obtained from the solution of the Klein--Gordon
  equation with a microscopic potential calculated within the $\Delta-$hole
  model~\cite{GarciaRecio:1989xa}, as employed
  in Refs.~\cite{AlvarezRuso:2007tt, AlvarezRuso:2007it, Amaro:2008hd}.  
To compute our results, we have used modified
  harmonic oscillator $\left [\rho(r) = \rho_0
  (1+a(r/R)^2)\exp(-(r/R)^2)\right]$ densities for carbon and oxygen
  and a two parameter Fermi distribution $\left[ \rho(r) = \rho_0
  /(1+\exp((r-c)/a_0))\right]$ for calcium. Density parameters:
  $R=1.692$ fm, $a= 1.082$ and $R=1.833$ fm, $a= 1.544$ for $^{12}$C
  and $^{16}$O; $c=3.51$ fm and $a_0=0.563$ fm for $^{40}$Ca.
 Data are from Refs.~\cite{Binon:1970ye} ($^{12}$C),
\cite{Albanese:1980np} ($^{16}$O), \cite{Gretillat:1981bq} ($^{40}$Ca
at $T_\pi=230$ MeV) and \cite{Marlow:1984ak} ($^{40}$Ca at $T_\pi=673$
MeV). }\label{fig:2IM}
\end{figure}

What we see in Fig.~\ref{fig:2IM} (solid lines), for all nuclei and pion
energies shown, is that both the size and the angular
dependence predicted by this model derived from the RS approach
strongly disagree with data. We have explored the available data for
other nuclei and pion energies and the outcome is similar.
Within this model, the size of the elastic cross section
could be largely affected by even small variations of $R_0$ (see
Eq.~(\ref{eq:fabs-pion})), and moreover it is not very clear why one
should use $\sigma_{inel}$ in the absorption factor, as we  discuss below.

In the case of coherent pion production induced by neutrinos one
expects distortion effects, accounted for by $F_{\rm abs}$, to be less
relevant and besides one should bear in mind that experimental
analyses do often adjust the size of the cross section with a free
parameter.  However, those neutrino analyses always rely on a
theoretical model for the outgoing pion angular distribution. As shown
in the figure, the RS derived model leads for low pion energies
 to $\pi$--nucleus differential cross
sections much flatter than experiment for all nuclei
and one would expect this will also be the case for coherent pion
production induced by low energy neutrinos. From this latter perspective, we
tentatively conclude that the RS model for neutrino--induced coherent pion
production, widely used in the literature, might
induce important uncertainties in the analysis of the neutrino
oscillation experiments for low energy incident neutrinos.

We discuss now the results that one obtains with the modifications to the RS 
model introduced in section~\ref{sec:im}. The pion-nucleus scattering 
would now read
\begin{equation}
\frac{d\sigma(\pi^0 {\cal N}_{gs} \to \pi^0 {\cal N}_{gs})}{dt} =
|F_{\cal A}^{distor-\pi}(\vec{q},\vec{k}_\pi)|^2 
\frac{d\sigma_{nsf}(\pi^0 N \to \pi^0
  N)}{dt}\,,\label{eq:piN}
\end{equation}
where 
\begin{eqnarray}
|F_{\cal A}^{distor-\pi}(\vec{q},\vec{k}_\pi)|^2 &\equiv& 
\Big | \int d^3\vec{r}\ e^{{\rm
i}\left(\vec{q}-\vec{k}_\pi\right)\cdot\vec{r}} 
\rho(\vec{r}\,)\Gamma^{\pi}(b) \Big |^2 \\
\Gamma^{\pi}(b) &=& \exp\left(
-\frac12 \sigma_{inel}\int_{-\infty}^{+\infty} dz' 
\rho\left( \sqrt{b^2+z^{\prime 2}} \right)\right)\,.
\end{eqnarray}
The profile $\Gamma^{\pi}$ does not depend on $z$ since 
both the incoming and outgoing pions should now be distorted. 

We show our predictions  (Eq.~(\ref{eq:piN})) in  Fig.~\ref{fig:2IM}
with dashed-lines.
As can be seen in the plots the two simple
corrections ($t-$dependence of the pion-nucleon cross section and a more
realistic description of the  pion distortion) 
 lead to a better description of the angular dependence,  
 improving enormously  previous RS based model results.

Finally, we would like to point out that it is not clear why one should use
$\sigma_{inel}$ in the distortion profile. Indeed, within the eikonal
approximation one should use the imaginary part of a realistic
pion-nucleus optical potential~\cite{Singh:2006bm}. This would eliminate
those pions that are absorbed, undergo quasielastic processes or
suffer inelastic reactions, taking into account Pauli blocking, Fermi
motion and other many--body effects to evaluate the corresponding
reaction probabilities. Such kind of sophisticated optical potential
is only available for low energy pions, up to the region of the
$\Delta (1232)$ resonance (see for instance
Refs.~\cite{GarciaRecio:1989xa,Nieves:1993ev,Nieves:1991ye}), and it
has been used in the microscopical approaches of
Refs.~\cite{Amaro:2008hd,Singh:2006bm, AlvarezRuso:2007tt,
AlvarezRuso:2007it}. Its extension to higher energies is a highly
non-trivial task. Here, we are not so much concerned with the size of
the cross sections but rather with the outgoing pion angular and energy
dependence. For this latter purpose, it might be
sufficient to use $\sigma_{inel}$ to compute the distortion profile.
 It is true that in this way, we do neither account
for pion absorption nor for pion quasielastic distortion (induced by
pion--nucleon elastic scattering). 
 However, this is partially compensated since the
use of $\sigma_{inel}$ leads to a larger distortion 
than it would be expected, if  the
inelastic processes were considered in the nuclear medium and Fermi
statistics was taken into account. The same discussion applies to the profile
function, $\Gamma(b,z)$, relevant for the neutrino coherent pion
production reaction, and defined in Eq.~(\ref{eq:gamma}).

\section{Neutrino-induced coherent $\pi^0$ production results}
\label{sec:re}

Once our framework has been satisfactorily tested against elastic
pion-nucleus differential cross section data, we give here results for
neutrino-induced coherent $\pi^0$ production. We work in the
neutrino-nucleus LAB frame. For the $x,y,t$ differential coherent
$\pi^0$ production cross section we use
\begin{eqnarray}
\frac{d\sigma}{dx\,dy\,dt}&=& \frac{G^2 M
E}{\pi^2}f_\pi^2 (1-y) \frac{H\left[1-|\cos\theta|\right]
}{(1-q^2/1\,{\rm GeV}^2)^2} |F_{\cal
  A}^{distor}(|\vec{q}\,|,|\vec{k}_\pi\,|, \vec{q}\cdot\vec{k}_\pi) |^2
 \frac{d\sigma_{nsf}(\pi^0 N \to \pi^0  N)}{dt}\Big|_{E_\pi=q^0}\,,
 \label{eq:sigma-xyt}
\end{eqnarray}
with $H[...]$ the step function and $\theta$  the center of mass 
pion  angle in the free pion--nucleon elastic reaction (see
Eq.~(\ref{eq:coscm})). 

That defines our model I.  We will also consider a second model (II),
where some kinematical corrections, recently proposed by Berger and
Sehgal~\cite{Berger:2008xs}, are incorporated. Those corrections, and
their degree of arbitrariness, were already discussed in
Section~\ref{sec:sf} and they amount to replace in
Eq.~(\ref{eq:sigma-xyt})
\begin{equation}
(1-y) \leftrightarrow \frac{q^0}{|\vec{q}\,|} u\,v =
\frac{q^0}{|\vec{q}\,|} \left (1-y + \frac{q^2}{4E^2} \right)\,.
\label{eq:berger} 
\end{equation}
 We also incorporate these corrections in the original RS model
performing the above replacement in Eq.~(\ref{eq:rsNucleon}). In what
follows, we will refer to this latter model as RS$^*$.  Although we
will look at the effect of the kinematical corrections proposed in
Ref.~\cite{Berger:2008xs}, we will not present results obtained from
the full approach followed in that reference. The reason being that in
Ref.~\cite{Berger:2008xs} the use of the experimental pion-nucleus
elastic cross section is advocated. This is not fully correct because
of the strong distortion of the incoming pion, implicit in the
experimental cross section data, that should not be taken into account
in neutrino-induced coherent pion production.

\subsection{$q^2-$distributions}
We readily obtain $d\sigma/dq^2$ from Eq.~(\ref{eq:sigma-xyt}) by
integrating  over $y$ and $t$  and performing the change of
variables $x \leftrightarrow q^2$. 
\begin{figure}
\begin{center}
\makebox[0pt]{\hspace{-15mm}
\includegraphics[width=10cm]{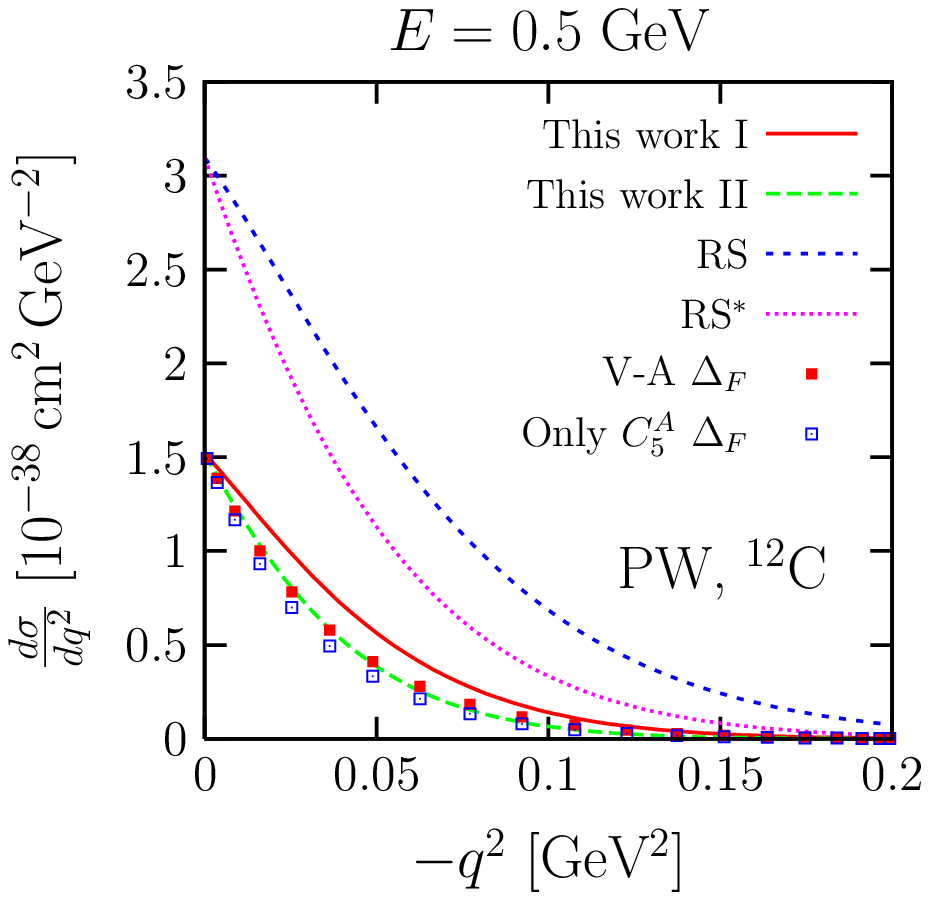}\hspace{-5mm}
\includegraphics[width=10cm]{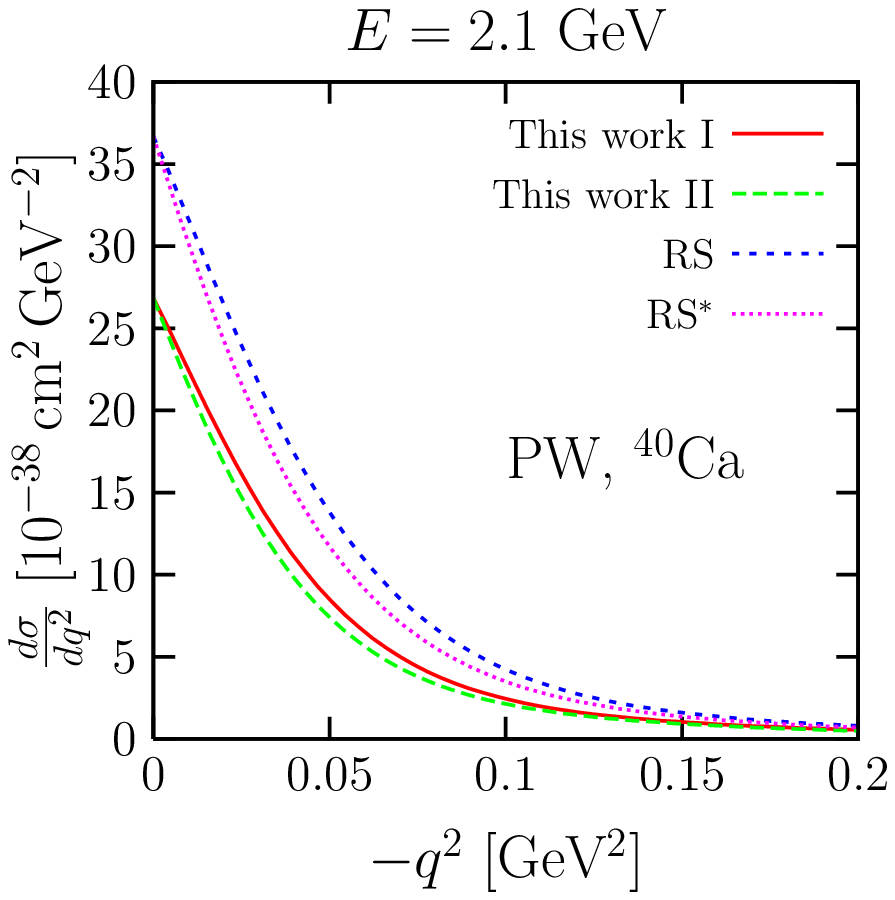}}
\end{center}
\caption{Plane Wave (no pion distortion) $q^2$ distributions obtained from models I
and II, together with those deduced from the RS and RS$^*$ models. In
addition for E=0.5 GeV in carbon, we also display results from  the
microscopic approach described in the text. Filled (open) squares
stand for the calculation with the full V--A (only with the $C_5^A$
contribution) $N\Delta$ current.}
\label{fig:figq2}
\end{figure}
In Fig.~\ref{fig:figq2}, we show results obtained without distortion of
the final pion. Eliminating distortion allows us to check, by
comparison with a microscopical calculation, the validity of the $t=0$
approximation used in the RS and RS$^*$ models.  With that aim, in
carbon and for E=0.5 GeV, we also display results from the microscopic model
of Ref.~\cite{Amaro:2008hd}.
At this energy,  neutrino-induced coherent pion production is dominated by the
excitation of the $\Delta(1232)$ resonance and its subsequent decay
into a $\pi^0 N$ pair. Thus, for simplicity, we have neglected the tiny
effects of the non-resonant background and the crossed$-\Delta$
term included in Ref.~\cite{Amaro:2008hd}. Besides, as all discussed models are 
based upon the free pion
nucleon cross section, nuclear medium effects that modify the $\Delta$
propagator have not been included.
We consider two cases, one that takes into account
the full $N\Delta$ weak current (V--A $\Delta_F$), and another that 
only includes the dominant axial term proportional to the $C_5^A$ form
factor\footnote{See for instance Eq.~(1) of
Ref.~\cite{Schreiner:1973ka} for a form factor decomposition of the
$N\Delta$ weak current.}  (Only $C_5^A\ \Delta_F$).  PCAC based models
rely on Eq.~(\ref{eq:pcac}) which relates, at $q^2=0$, a weak matrix
element with a purely hadronic one.  In the case of the $N\Delta$
transition, the only term of the divergence of the axial current that
survives at $q^2=0$ is that proportional to the form factor
$C_5^A$. Thus, to make meaningful the comparison between the PCAC
based models examined here and the microscopic approach including
explicitly a $\Delta$, requires fixing $C_5^A(0)$ to the value
predicted by the non--diagonal Goldberger--Treiman relation
($C_5^A(0)\equiv\sqrt{\frac23}f_\pi\frac{f^*}{m_\pi}=1.2$, with the
$\pi N \Delta$ coupling $f^* = 2.2$ fixed to the $\Delta$ width).
  We see that I and II calculations
are in reasonable agreement with the microscopical model, whereas the
RS and RS$^*$ ones predict much larger and wider differential cross
sections. As mentioned above, this is mainly due to the $t=0$
approximation assumed in those models\footnote{The used nuclear
form-factors are also different (see text after
Eqs.~(\ref{eq:rsNucleon}) and (\ref{eq:rsxxx})), but their effect
can not account for the large differences observed.  As a matter of
example in $^{12}$C and for E=0.5 GeV, the plane wave RS model gives
an integrated cross section (in units of $10^{-38}$ cm$^2$) of 0.206,
while our model I predicts a value of 0.070 in the same units. If we
use model I with the nuclear form factor proposed in the original RS
model, we find an integrated cross section of 0.089 $10^{-38}$
cm$^2$.}. The effects due to the $t=0$ approximation decrease with the
neutrino energy and atomic number, but  they are still important
for E=2.1 GeV in a medium sized nucleus like calcium.  At $q^2=0$, and
in the absence of distortion, our model should essentially match the
microscopical calculation (as it was discussed in Sect.~ IVB-1 of
Ref.~\cite{Amaro:2008hd}), and indeed the two models nicely agree at
this kinematical point.

 Far from $q^2=0$, PCAC based models just consider the axial part of
the current, that given the tiny effect of the vector part in the
microscopical model (as can be seen by comparing ``V--A $\Delta_F$'' and
``Only $C_5^A\ \Delta_F$'' microscopic results), it turns out to be an
excellent approximation here. Our model II describes the ``Only
$C_5^A\ \Delta_F$'' results better than model I, which make us conclude
that the kinematical corrections recently proposed
by Berger and Sehgal~\cite{Berger:2008xs} are sound.

The small differences at finite $q^2$ can be due to some discrepancies
between the experimental pion-nucleon cross section and that deduced
from the $\Delta$ mechanism alone (with $f^*$ fixed from the $\Delta$
width), to offshellness ($q^2\ne m_\pi^2$) effects, or to different
treatments of the nucleon dynamics inside the nuclear medium. Besides,
there are other effects that vanish at $q^2=0$ and that are present in
this case. For instance,  in the microscopical model  the
$\sigma_S$ structure function (Eq.~(\ref{eq:sigma_S})) is
not proportional to $q^\mu q^\nu H_{\mu\nu}$ and hence $\sigma_S$
cannot be exactly related to the pion-nucleus cross section for
$q^2\ne 0$ values. Similarly, corrections due to the $\sigma_R$,
$\sigma_L$ structure functions in Eq.~(\ref{eq:sigmarsl}) are not
taken into account within the PCAC approximation. Other sources of
discrepancies come from the adopted $q^2$ behaviour of $C_5^A(q^2)$~\cite{Amaro:2008hd},
which gives rise to a faster $q^2-$decrease than that provided by the
propagator term $1/(1-q^2/m_A^2)^2$ included in the PCAC based models.
Nonetheless, at least for this observable, the effect of all these
differences is minor and model II gives a good approximation to the
microscopical calculation\footnote{A word of caution must be said
here.  The basic PCAC formula in Eq.~(\ref{eq:pcac}) might suffer from
corrections. For instance, in the $\Delta$ region, a value for
$C_5^A(0)$ significantly smaller (0.87) than 1.2 was used in
Ref.~\cite{Amaro:2008hd}. This reduced coupling would lead to cross
sections around a factor of two [$(1.2/0.87)^2$] smaller than those
presented in Fig.~\ref{fig:figq2}. This violation of the off-diagonal
Goldberger-Treiman relation by about 30\% was proposed in
Ref.~\cite{Hernandez:2007qq} after fitting the Argonne bubble chamber
$\nu_\mu + p \to \mu^- \pi^+ p$ cross section
data~\cite{Radecky:1981fn}, including a non-resonant
background. Non-resonant background terms, though
important at the nucleon level, turn out to be negligible in the
neutrino coherent pion process~\cite{Amaro:2008hd}. The lattice QCD
results shown in Figure 4 of Ref.~\cite{Alexandrou:2006mc} might also
support a value for $C_5^A(0)$ smaller than
$\sqrt{\frac23}f_\pi\frac{f^*}{m_\pi}$.  This is indeed a serious
problem that deserves a joined theoretical and experimental effort to
sort it out.}.

\begin{figure}[ht]
\begin{center}
\makebox[0pt]{\hspace{-15mm}
\includegraphics[width=10cm]{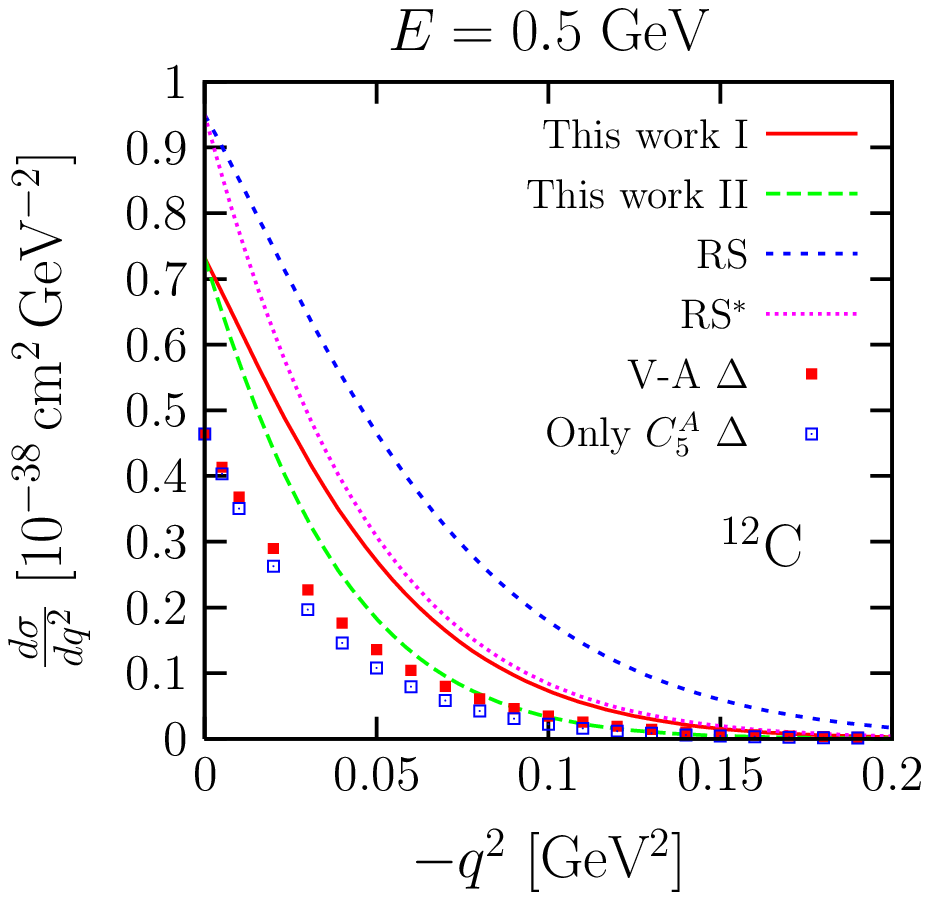}\hspace{-5mm}
\includegraphics[width=10cm]{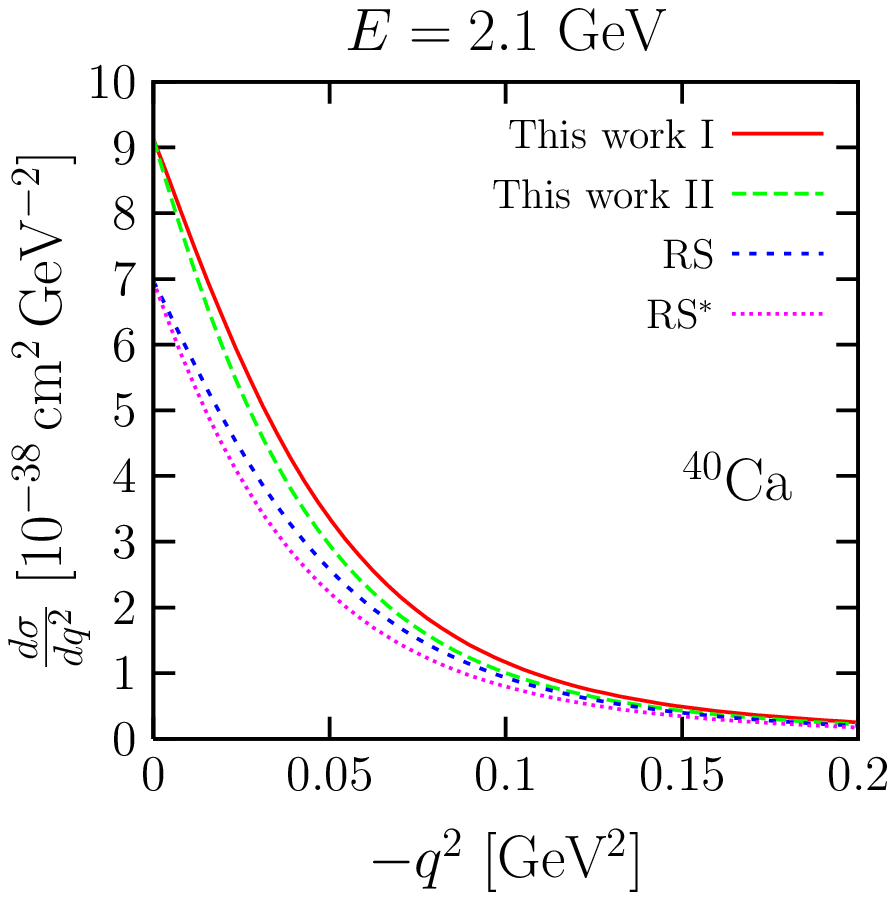}}
\end{center}
\caption{Full $q^2$ distributions (including distortion of the
outgoing pion).  For E=0.5 GeV in carbon we also display results from
the microscopic approach of Ref.~\cite{Amaro:2008hd} with
$C_5^A(0)=1.2$. Filled (open) squares stand for the calculation with
the full V--A (only with the $C_5^A$ contribution) $N\Delta$ current.}
\label{fig:figq2bis}
\end{figure}

Now we look at the results with distortion of the final pion that we
show in Fig.~\ref{fig:figq2bis}.  
In this case, for comparison, we use the model based on the microscopic approach
developed in Refs.~\cite{AlvarezRuso:2007it, Amaro:2008hd}
including both distortion and the
nuclear medium effects affecting the $\Delta(1232)$. Again, and for simplicity, 
we have neglected the tiny
effect of the non-resonant background and the crossed$-\Delta$
term included in Ref.~\cite{Amaro:2008hd}. 

Distortion reduction is
significantly larger in the RS and RS$^*$ models than in our
scheme. That diminishes the differences in the full calculation in
carbon, and explains the reverse pattern observed in calcium. 
Note, however, that the pion distortion in the RS and RS$^*$
schemes is too strong. Indeed, it leads to forward pion-nucleus
elastic cross sections smaller than both data and those deduced within
our framework, as can be appreciated in Fig.~\ref{fig:2IM}. In spite  of
that, both the RS and RS* models still overestimate the
microscopical calculation in carbon by a large factor. 

On the other hand, the situation is somehow puzzling when we compare
microscopical results with those of model I or II. As discussed above, within
the PW approximation both types of approach nicely agree, but distorted
results notably disagree. This is despite the fact that the
Klein--Gordon $\Delta-$hole model, in which the
microscopical results depicted in Fig.~\ref{fig:figq2bis} are based, leads to
elastic pion nucleus cross sections of similar quality as those
deduced from the eikonal framework developed here, and that our model
I and II use (see for instance the comparison performed in the right
top panel of Fig.~\ref{fig:2IM} for a typical pion energy). This
contradiction deserves some discussion, and we believe it is related
to the inadequacy of the use of eikonal distortion for low energy
pions and to the modifications of the $\Delta$ resonance 
properties in the nuclear medium.

Describing correctly the elastic pion--nucleus cross section is a
necessary condition that should satisfy any model aiming at a proper
description of the coherent pion production process induced by
neutrinos. However, that constraint is not sufficient, and the
situation in Fig.~\ref{fig:figq2bis}  is an example of that.

One of the major sources of differences between our model of
Eq.~(\ref{eq:piN}) and the microscopic calculation is due to in-medium
modifications of the free pion-nucleon cross section in the latter
case (see for instance Ref.~\cite{Amaro:2008hd}). Within the eikonal
model assumed here to construct the elastic pion-nucleus cross
section, there are two competing factors: the pion--nucleon elastic
non spin flip $\sigma_{\rm nsf}$ cross section and the pion
distortion, controlled by $\sigma_{\rm inel}$. As a consequence, the
pion--nucleus scattering amplitude does not linearly depend on the
corresponding pion--nucleon one. That explains why different
approaches for the in medium pion--nucleon amplitude could 
lead to similar pion-nucleus cross sections. For instance, let us
suppose that in medium modifications reduce (enhance) the pion--nucleon cross
sections. We have that pion distortion effects
become less (more) important with decreasing (increasing) 
$\sigma_{\rm inel}$ and this reduced (enhanced)
suppression can compensate a decrease (an enhancement) 
in $\sigma_{\rm nsf}$. The
outcome is that you can predict similar pion-nucleus cross sections
starting from quite different pion-nucleon cross sections.
It seems that induced differences in the pion distortion compensate
the changes due to the medium in the elastic pion-nucleon cross
section in pion-nucleus scattering, but not in neutrino induced
coherent pion production where distortion becomes smaller and 
affects only the outgoing pion.

For low energy pions, of interest in MiniBooNE and T2K, we believe that
the distorted results from the microscopical models of 
Refs. ~\cite{AlvarezRuso:2007it,Amaro:2008hd} are more realistic that
those based on the eikonal approximation presented here, where the
pion-nucleon interaction is not modified in the medium and a simple
model based on $\sigma_{inel}$ is used to distort the outgoing pion
waves. The framework of Ref.~\cite{Singh:2006bm}, despite the use
of the eikonal approximation, greatly overcomes these shortcomings,
since there, the modification of the $\Delta$ properties in the medium
are taken into account, and the imaginary part of a realistic
pion-nucleus optical potential is used to distort the outgoing
pion. The model becomes as complicated as that of
Refs.~\cite{AlvarezRuso:2007it,Amaro:2008hd}, as difficult as this
latter one to extrapolate at higher pion energies, and  it is
still less reliable, since multiple scattering is not taken into
account within the eikonal approximation as accurately as by solving
the Klein Gordon equation~\cite{AlvarezRuso:2007it,Amaro:2008hd}.

\subsection {Pion distributions}

For NC driven processes, the $q^2$ distribution can not be easily
measured because of the obvious difficulty in detecting the outgoing
neutrino or the nucleus recoil. Thus, it is of great interest to examine pion
distributions.  

\subsubsection{Pion energy distributions}

\begin{figure}[h]
\begin{center}
\makebox[0pt]{\hspace{-25mm}
\includegraphics[width=10cm]{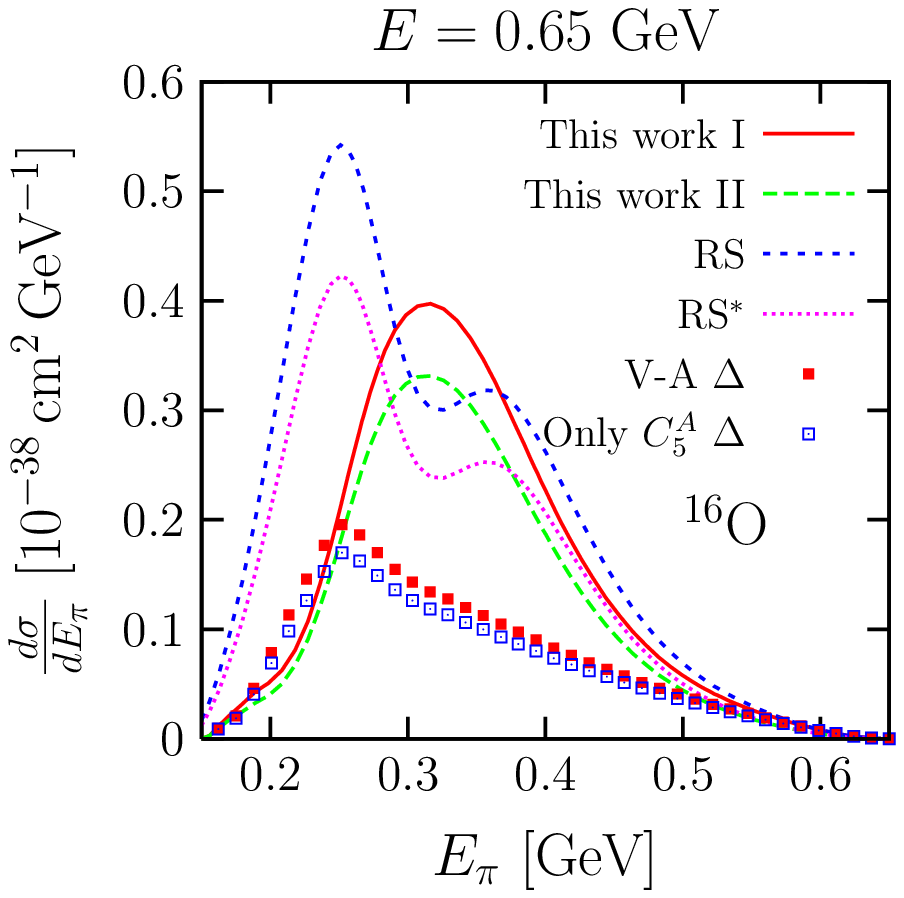}\hspace{-5mm}
\includegraphics[width=10cm]{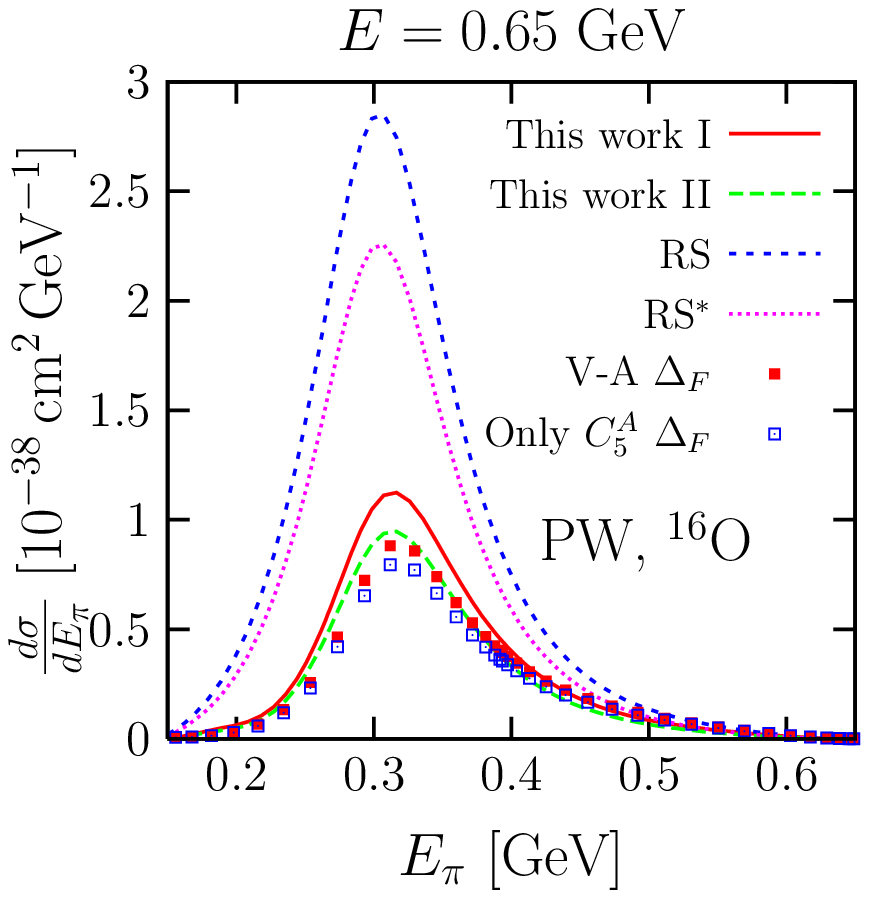}}
\end{center}
\caption{Distorted (left panel) and plane wave (right panel)
  $d\sigma/dE_\pi$ results in oxygen obtained from our models I and
  II, together with those deduced from the original RS and the RS$^*$
  models. The incoming neutrino energy is 0.65 GeV. In addition, we
  also display results from the microscopic approach of
  Ref.~\cite{Amaro:2008hd} with $C_5^A(0)=1.2$. Filled (open) squares
  stand for the calculation with the full V--A (only with the $C_5^A$
  contribution) $N\Delta$ current. }
\label{fig:figepi}
\end{figure}

In Fig.~\ref{fig:figepi}, we show the outgoing pion spectrum from
oxygen, with and without distortion effects, predicted by the different
models considered in the previous subsection. The incoming neutrino
energy is 0.65 GeV. The integrated cross sections are compiled in Table
\ref{tab:oxy}.  

We see once more that, for the case without distortion, our model II
has a reasonable agreement with the ``V--A $\Delta_F$'' and ``Only
$C_5^A \Delta_F$'' microscopic calculations. Not only the shapes, which
peak at around $E_\pi =320$\,MeV, but also the integrated cross
sections are similar (see the table).  Again the RS and RS$^*$ models
give much larger results and the position of the peak is slightly
displaced towards lower energies.

In the distortion case we find totally different behaviours. Results
obtained with our models I and II show a shape similar to the case
without distortion, while the RS and RS$^*$ calculations show a double
peak structure. In the microscopical calculation a single peak
appears, but now at a lower pion energy $E_\pi\approx 260$\,MeV. This
change in the position of the peak in the microscopical results can be
traced to the modification of the properties of the $\Delta$ in the
medium and to nonlocal effects in the amplitude (see for instance
Fig.~2a in Ref.~\cite{Amaro:2008hd}). These non-localities reflect the
fact that the pion three-momentum is only well defined asymptotically
when the pion-nucleus potential vanishes~\cite{Amaro:2008hd}. These
effects are neither present in any PCAC based model where one uses
the free-space pion nucleon cross sections, nor in any microscopic
calculation with plane waves. 

At low and intermediate neutrino energies, once the distortion is
considered, we must again conclude  that the microscopic model
predictions are most reliable than those deduced from the PCAC based
model examined here. The original RS predictions have little reliability
 and the hope is that at higher energies, where we expect
 nuclear medium modifications of the elementary pion--nucleon
interaction  to be less relevant, the model I and II could become better 
suited.

\begin{table}
 \begin{center} 
\begin{tabular}{ccccccc}\hline\tstrut
 & & & \multicolumn{2}{c}{Microscopical}\\\cline{4-5}\tstrut
& ~This work I~ &~ This work II ~& ~~V--A $\Delta$~~~ & ~~~Only $C_5^A\,\, \Delta$~~~ & 
~RS & ~~RS$^*$~ \\\hline \tstrut
PW & 0.152 & 0.127 & 0.117 & 0.106 & 0.389 & 0.305  \\
DW & 0.072 & 0.060 & 0.037 & 0.033 & 0.100 & 0.078  \\
\hline 
\end{tabular}
 \end{center} 
\caption{Integrated cross sections (in units of $10^{-38}$ cm$^2$)
  corresponding to the different curves displayed in
  Fig.\protect\ref{fig:figepi}. PW and DW stand for plane wave and
  distorted results, respectively. }
\label{tab:oxy} 
\end{table} 

\subsubsection{Pion angular distributions}

The starting point of the PCAC based models is the $x,y,t$
differential cross section. Firstly, we would like to stress that for
$q^2\ne 0$, the
knowledge of $\frac{d\sigma}{dx\,dy\,dt}$ is not sufficient to compute
the angular distribution of the outgoing pion with respect to the
direction of the incoming neutrino.
Let us  take here as z-axis the direction of the 
incoming neutrino. The pion and transferred momenta are given by
\begin{eqnarray}
 \vec{k}_\pi&=&|\vec{k}_\pi|\left (\sin\theta_\pi\cos\phi_\pi,
 \sin\theta_\pi\sin\phi_\pi, \cos\theta_\pi   \right)\,, \\
 \vec{q}&=& \left(-|\vec{k}'|\sin\theta',0, |\vec{k}\,|
 -|\vec{k}'|\cos\theta'
\right)\,,
\end{eqnarray}
with $\theta'$ the outgoing neutrino scattering angle, $\theta_\pi$
the angle between the incoming neutrino and the outgoing pion and
$\phi_\pi$ the azimuthal pion angle in this frame. By means of a
rotation that takes $\vec q$ along the positive $z-$axis one can
easily obtain
\begin{equation}
\cos\theta_\pi = \hat{k}\cdot\hat{k}_\pi =
  \frac{|\vec{k}'|}{|\vec{q}\,|}\sin\theta'\sin\theta_{k_\pi
  q}\cos\phi_{k_\pi q}+
  \frac{|\vec{k}\,|-|\vec{k}'|\cos\theta'}{|\vec{q}\,|}\cos\theta_{k_\pi
  q} \,,
\end{equation}
where $\theta_{k_\pi q}$ and $\phi_{k_\pi q}$ are the pion polar and
azimuthal angles in that frame where the positive $z-$axis is taken in the 
direction of $\vec{q}$. 

The incoming neutrino energy and the variables $x$ and $y$ determine
$|\vec{k}'|$, $|\vec{q}\,|$ and $\theta'$, while, within the $E_\pi
=q^0$ approximation, $t$ fixes $\theta_{k_\pi q}$ [$t=
-\vec{q}^{\,\,2}-\vec{k}^2_\pi+2|\vec{q}\,\,| |\vec{k}_\pi|
\cos\theta_{k_\pi q}$ ], remaining $\phi_{k_\pi q}$
undetermined. Thus, as stated before, the knowledge of
$d\sigma/dx\,dy\,dt$ is not sufficient to compute the angular
distribution of the outgoing pion with respect to the direction of the
incoming neutrino, and it would be necessary to know
$\frac{d\sigma}{dx\,dy\,dt\,d\phi_{k_\pi q}}$. Only for
$q^2=0$, $\theta_{k_\pi q} = \theta_\pi$, since $\theta'=0$ (both the
outgoing neutrino and the momentum transfer go along the incoming
neutrino direction) and $\frac{d\sigma}{dx\,dy\,dt}\big|_{q^2=0}$
determines $\frac{d\sigma}{dx\,dy\,d\cos\theta_\pi}\big|_{q^2=0}$.

In PCAC based models it is assumed,
\begin{equation}
\frac{d\sigma}{dx\,dy\,dt\,d\phi_{k_\pi q}} 
= \frac{1}{2\pi} \frac{d\sigma}{dx\,dy\,dt}
= \frac{1}{2\pi}\int d\phi_{k_\pi q}\, 
\frac{d\sigma}{dx\,dy\,dt\,d\phi_{k_\pi q}}\,,
\label{eq:ang-approx}
\end{equation}
which leads to~\cite{Rein:1982pf}
\begin{eqnarray}
\frac{d\sigma}{dE_\pi d\eta\, d\cos\theta'd\phi_\pi} &=& 
\frac{E-E_\pi}{M E_\pi^2} \frac{|\vec{k}_\pi|\,|\vec{q}\,|}{\pi}
H\left[1-|\cos\theta_\pi|\right]
 \frac{d\sigma}{dx\,dy\,dt} \,,
\label{eq:eta}
\end{eqnarray}
where $\eta=E_\pi(1-\cos\theta_\pi)$ is the variable proposed by the
MiniBooNE Collaboration in its recent analysis of coherent $\pi^0$
production of Ref.~\cite{AguilarArevalo:2008xs}.  

For non-vanishing $q^2$ values, Eq.~(\ref{eq:ang-approx}) is
incorrect, and therefore Eq.~(\ref{eq:eta}) is wrong as well. The
problem arises because $\frac{d\sigma}{dx\,dy\,dt\,d\phi_{k_\pi q}}$
depends in general on $\phi_{k_\pi q}$, through the $\frac{d{\cal
A}}{dt\,d\phi_{k_\pi q}}$ term in Eq.~(\ref{eq:dsigmarsl}).  When
$q^2$ is zero that term does not contribute and thus
Eqs.~(\ref{eq:ang-approx}) and (\ref{eq:eta}) are correct only in this
limit.

\begin{figure}[ht]
\begin{center}
\makebox[0pt]{\hspace{-15mm}
\includegraphics[width=10cm]{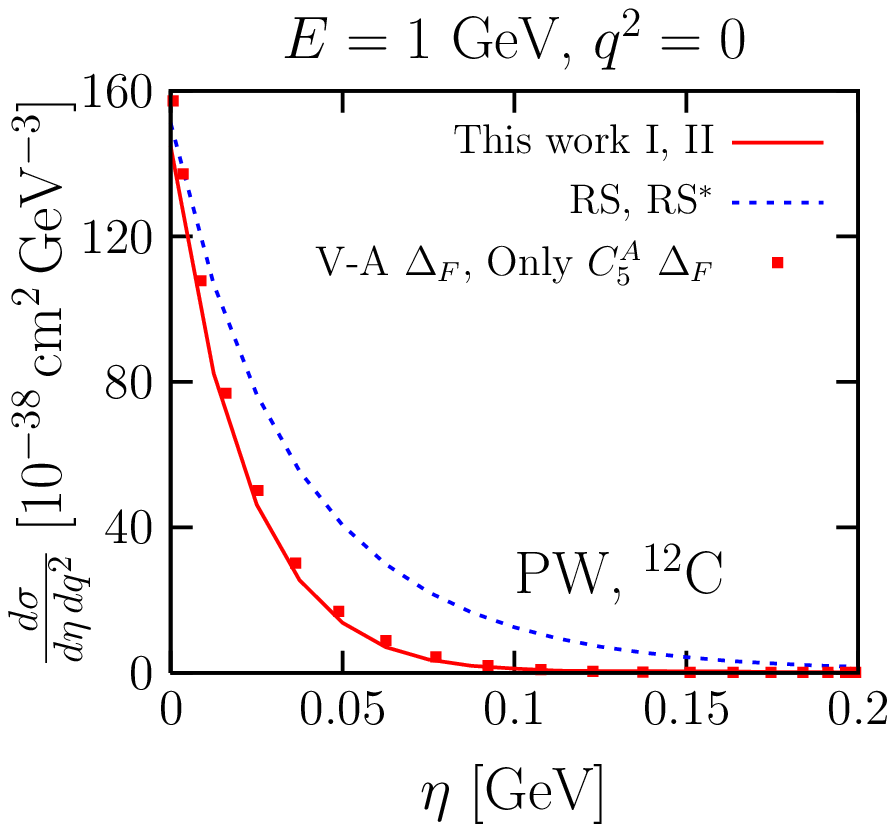}\hspace{-5mm}
\includegraphics[width=10cm]{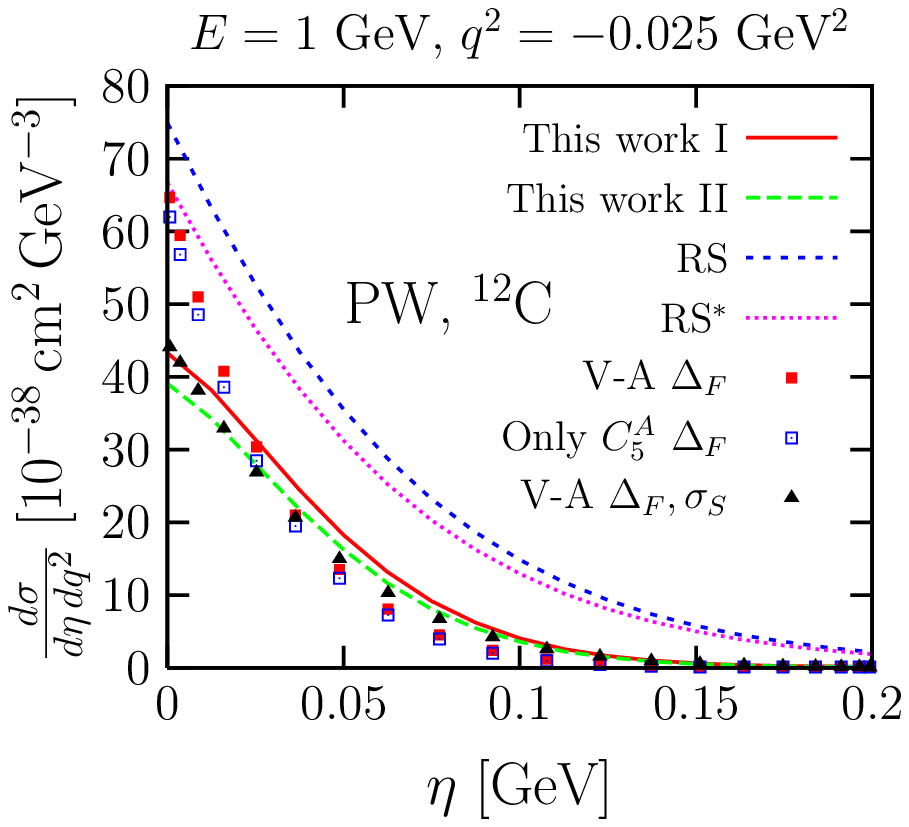}}\\[0.5cm]
\makebox[0pt]{\hspace{-15mm}\includegraphics[width=10cm]{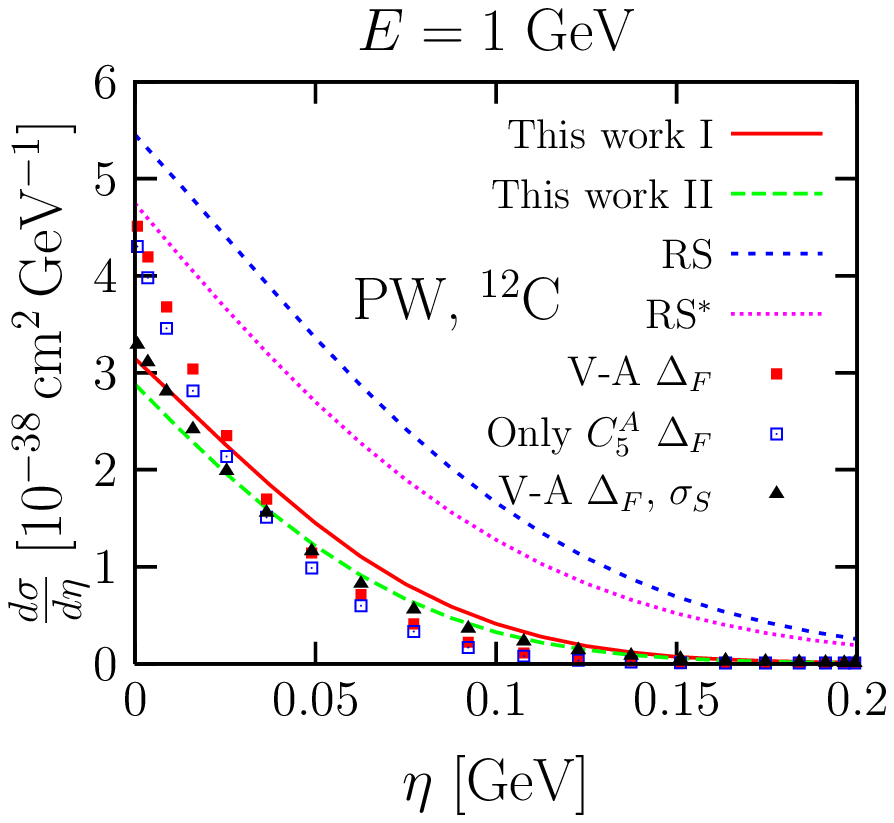}}
\end{center}
\caption{Undistorted $d\sigma/d\eta dq^2$ (top panels) and
$d\sigma/d\eta$ (bottom panel) differential cross sections obtained
from  models I, II,  RS and RS$^*$ and the microscopic
approach of Ref.~\cite{Amaro:2008hd} with $C_5^A(0)=1.2$. Filled
(open) squares stand for the calculation with the full V--A (only with
the $C_5^A$ contribution) $N\Delta$ current.  In the top-right and
bottom panels we also show results (triangles) for the ``V--A $\Delta_F$''
microscopic calculation including only the $\sigma_S$ contribution.}
\label{fig:figeta}
\end{figure}

Let us start by looking at undistorted results in
Fig.~\ref{fig:figeta}. In the two upper panels we show
$\frac{d\sigma}{d\eta\,dq^2}$ differential cross sections for two
different $q^2$ values, while in the lower panels we show $q^2$
integrated distributions. The nucleus is carbon and the incident
neutrino energy is $E=1$\,GeV.  For $q^2=0$ (top-left panel) we find a
very nice agreement between our models I and II calculation and the
microscopic model results for $\frac{d\sigma}{d\eta\,dq^2}$. On the
other hand, RS and RS$^*$ models give rise to  significantly flatter
distributions. In the top-right panel, we show results for $q^2=-0.25$\,
GeV. We find now a disagreement of the models I and II with the microscopic calculation. This
disagreement goes over to the $q^2$ integrated $\frac{d\sigma}{d\eta}$
differential cross section shown in the bottom panel. The microscopic
calculation is much more peaked at $\eta=0$ than either models I
and II or the RS and RS$^*$ models, the latter two showing the
flattest distributions.  The reason for the discrepancy between the
microscopic model and the PCAC based ones is due to two approximations
made in the latter models. The first approximation made in PCAC based
models is the one encoded in Eq.~(\ref{eq:ang-approx}) which leads to
Eq.~(\ref{eq:eta}) instead of the correct expression
\begin{eqnarray}
\frac{d\sigma}{dE_\pi d\eta\, d\cos\theta'd\phi_\pi} &=& 
2\frac{E-E_\pi}{M E_\pi^2} {|\vec{k}_\pi|\,|\vec{q}\,|}
H\left[1-|\cos\theta_\pi|\right]
 \frac{d\sigma}{dx\,dy\,dt\,d\phi_{k_\pi q}} 
\label{eq:etabis}
\end{eqnarray}
As mentioned above, the approximation in Eq.~(\ref{eq:ang-approx}) amounts to
neglect the $\frac{d{\cal A}}{dt\,d\phi_{k_\pi q}}$ term in
Eq.~(\ref{eq:dsigmarsl}), term that the microscopic calculation
properly takes into account. 

Besides, PCAC based models only consider the $\sigma_S$ contribution
in Eq.~(\ref{eq:dsigmarsl}) while the microscopic one also takes into
account the contributions from $\sigma_R$ and $\sigma_L$. To see the
effects associated to these neglected terms we also show in the top-right
and bottom panels the ``V--A $\Delta_F$'' microscopic calculation
considering only the $\sigma_S$ contribution.  What we see is that the
differential cross sections become much flatter than before, being now
in a nice agreement with our model II calculation. The fact that the
RS and RS$^*$ models give the flattest distributions is explained by
the further $t=0$ approximation used there.

The conclusion is that in PCAC models $\sigma_R$, $\sigma_L$ and
${\cal A}$ structure functions in Eq.~(\ref{eq:dsigmarsl}) cannot be
taken into account. The omission of these contributions leads to
significantly flatter $d\sigma/d\eta$ differential cross sections than
those predicted by a microscopic calculation. This affects both our
models I and II and the RS and RS$^*$ models. In the latter two cases
the $t=0$ approximation enhances this behaviour.

\begin{figure}[ht]
\begin{center}
\makebox[0pt]{\hspace{-15mm}
\includegraphics[width=10cm]{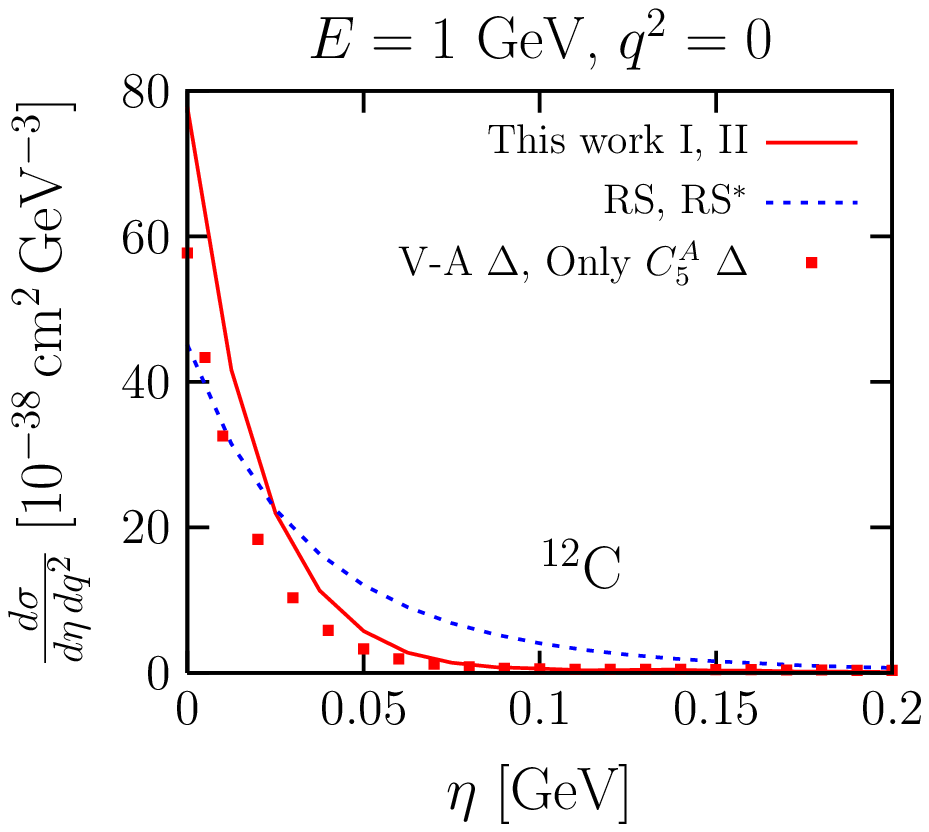}\hspace{-5mm}
\includegraphics[width=10cm]{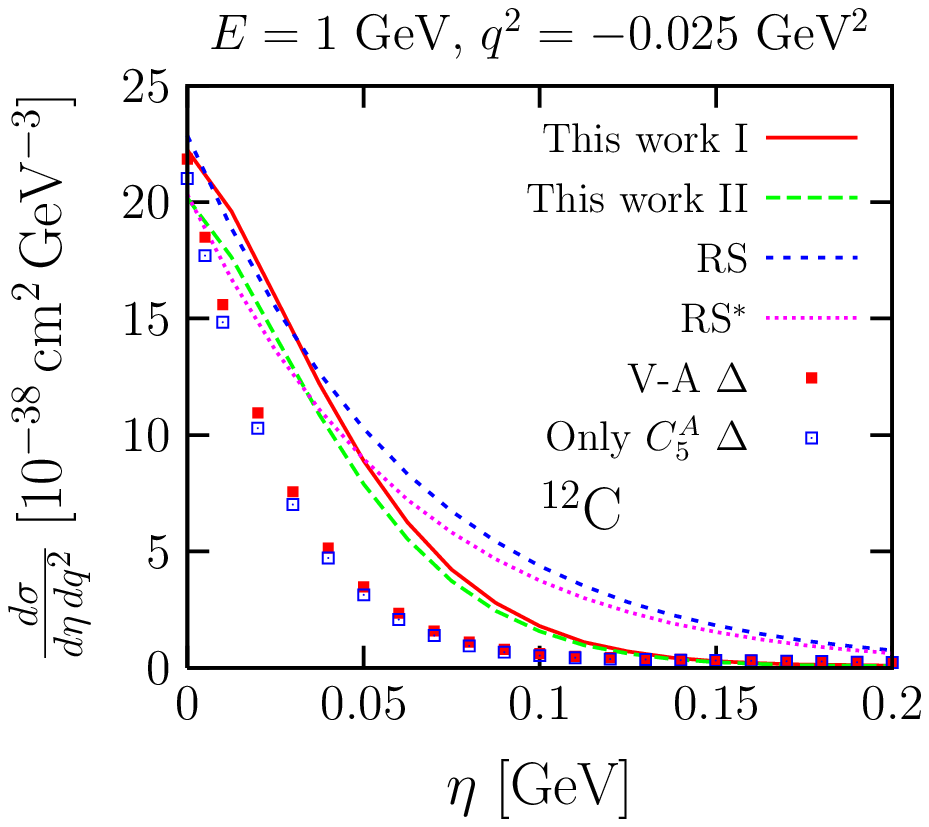}}\\[0.5cm]
\makebox[0pt]{\hspace{-15mm}\includegraphics[width=10cm]{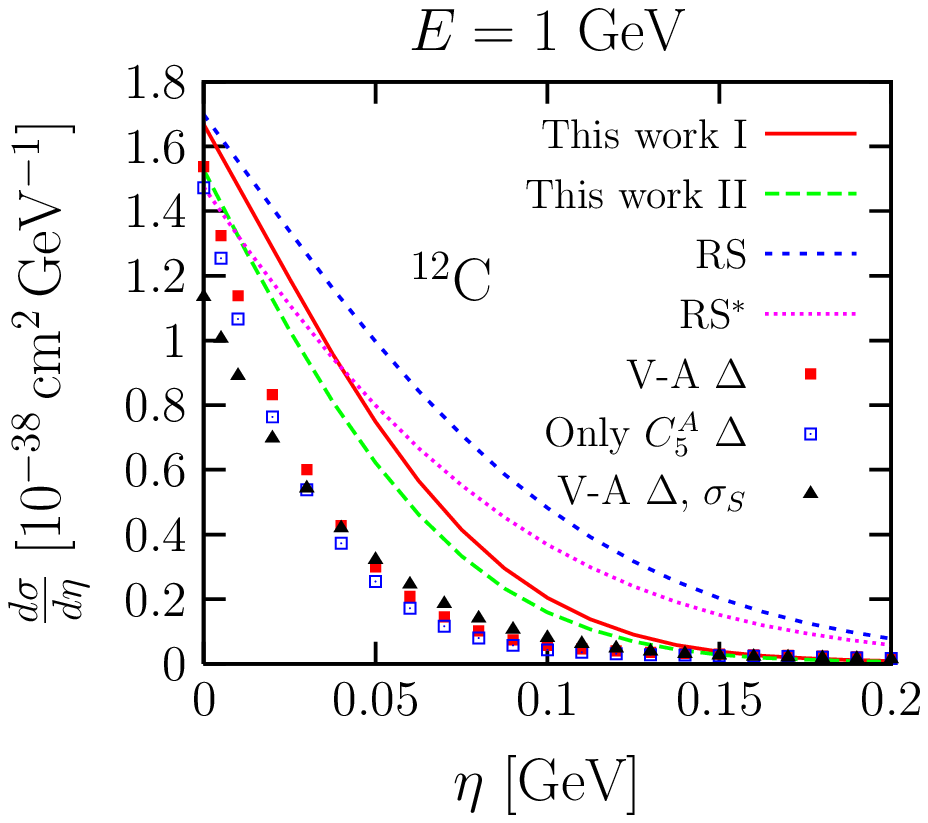}}
\end{center}
\caption{Same as Fig.~\ref{fig:figeta} including distortion.}
\label{fig:figeta2}
\end{figure}

This is also true for distorted results that are shown in
Fig.~\ref{fig:figeta2}. Again the microscopic calculation is more
peaked close to $\eta=0$, and the RS and RS$^*$ results show the
flattest distributions due to the $t=0$ approximation.
In any case, model RS$^*$ and even more model II represent a major 
improvement as compared to the original RS model for this observable.

These findings have immediate consequences for the results published
by the MiniBooNE Collaboration in its recent analysis of coherent
$\pi^0$ production of Ref.~\cite{AguilarArevalo:2008xs}, which rely on
the RS model. The distribution for $E_\pi
(1-\cos\theta_\pi)$ given there should
be definitely much narrower and much more peaked around zero, and
thus it might improve the description of the first bin value in
Fig.~3b of this reference. This was already suggested in
Ref.~\cite{Amaro:2008hd}.

\section{Summary and conclusions}
\label{sec:su}

We have critically reviewed the commonly used Rein--Sehgal model for
NC neutrino coherent pion production~\cite{Rein:1982pf}. We have
unambiguously pointed out the main deficiencies of this model, which
induce important uncertainties  for pions of relatively low energy,
as those of relevance in the MiniBooNE and T2K experiments. Among
others, the more relevant ones are:
\begin{enumerate}

\item The $t$ dependence of the coherent production is fully ascribed to the
  nuclear form factor $F_{\cal A}(t)$, while further and significant
  $t-$dependences induced by the 
  pion--nucleon interaction are ignored (see
  Fig.~\ref{fig:1NM}).
 
 The recent works of Refs.~\cite{Berger:2008xs,Paschos:2009ag} try to
 overcome this problem by using experimental information on the
 $t-$dependence of the elastic pion-nucleus cross section.  Apart from
 the obvious limitation coming from the lack of experimental data for
 many pion energies and nuclei, this might not be appropriate
 either for energies where the process is dominated by the weak
 excitation of the $\Delta(1232)$ resonance. As discussed in Sect.~\ref{sec:sf}
 there might be a non-trivial off-shell behaviour
 for $\sigma_{\pi^0{\cal N}}$~\cite{Bell:1964eu}.
In a microscopic approach one would argue that because of the
 strong distortion of the incoming pion in the on-shell elastic pion nucleus
 process at  energies in the $\Delta(1232)$ resonance region, one cannot 
 directly relate its amplitude
 to the pion production induced by a weak current. That takes us
 naturally to the second caveat.

\item The treatment of the outgoing pion distortion within the
  original RS model is quite poor, and it turns out to be certainly
  insufficient for resonant pions. An unquestionable evidence for
  that can be seen in Fig.~\ref{fig:2IM}, where it is shown that both the
  size and the angular dependence predicted by a model derived from
  the RS approach strongly disagree with the elastic pion--nucleus
  differential cross section data.

\item Far from the $q^2= 0$ kinematical point, any PCAC based model,
  and in particular the RS one, cannot be used to determine the
  angular distribution of the outgoing pion with respect to the
  direction of the incoming neutrino (see right top and bottom panels
  of Fig.~\ref{fig:figeta}).  Terms that vanish at $q^2=0$, and that
  are not considered in PCAC based models, provide much more forward
  peaked outgoing pion--incoming neutrino angular distributions.
  
PCAC models can only determine the distribution on
  the angle formed by the pion and the lepton transferred momentum,
  $\vec{q}$.  Experimentally, one can have access to this latter
  differential cross section in the case of CC driven processes, but
  not when the reaction takes place through the weak neutrino NC.

\end{enumerate}

We address the first of these deficiencies, and we succeed to
improve the original RS model by incorporating the $t-$dependence of
the pion-nucleon cross section. 

However, there is not an easy solution for the other two caveats at low
and intermediate pion energies.  

We have tried to improve the treatment of the distortion of the
outgoing pion, while keeping the model still reasonably
simple. Although  we have managed to describe the pion--nucleus elastic
cross section, we still find significant discrepancies in the case of
neutrino induced processes when we compare with the accurate
microscopical model of Refs.~\cite{AlvarezRuso:2007it,Amaro:2008hd}
(see Figs.~\ref{fig:figq2bis} and~\ref{fig:figepi}).  This illustrates
a further interesting point: describing correctly the elastic
pion--nucleus cross section is a necessary condition that should
satisfy any model aiming at a  proper description of the coherent pion
production process induced by neutrinos, but however it is not a
sufficient one. Discrepancies are due to the modification of the
elementary processes when they take place inside the nuclear medium,
 and to the highly non-linear character of the strong
driven processes.

 Altogether, we have only been  able to quantify the
unavoidable systematic error associated to these two last caveats for
low and intermediate energy pions by comparing with the 
model of Refs.~\cite{AlvarezRuso:2007it,Amaro:2008hd}. 

 All its limitations notwithstanding, the improved models I and II give a much
better description of low and intermediate energies coherent pion
production than the widely used RS model.  Of course, as
the neutrino energy increases, the $q^2=0$ kinematics becomes much
more dominant, and on the other hand we expect nuclear medium
modifications of the elementary pion--nucleon interaction to be less
relevant. Under these circumstances, our improved models become
even more appropriate. Nevertheless, we should point out the
existence of a real problem: for neutrino energies in the region 1 to
2 GeV, there does not exist a reliable model to describe the coherent
pion production process. This is because, these neutrino energies are
not large enough for PCAC models to properly work, and they are
certainly in the limit of applicability of microscopical models  which only
include the dynamics of the $\Delta(1232)-$resonance.

\begin{acknowledgments}
  This research was supported by DGI and FEDER funds, under contracts
  FIS2008-01143/FIS, FIS2006-03438, FPA2007-65748, and the Spanish
  Consolider-Ingenio 2010 Programme CPAN (CSD2007-00042), by Junta de
  Castilla y Le\'on under contracts SA016A07 and GR12, and it is part
  of the European Community-Research Infrastructure Integrating
  Activity ``Study of Strongly Interacting Matter'' (acronym
  HadronPhysics2, Grant Agreement n. 227431) under the Seventh
  Framework Programme of EU.
\end{acknowledgments}

\end{document}